\documentclass[onecolumn,aps,pre]{revtex4}
\usepackage{graphicx}
\usepackage[]{epsfig}
\usepackage{amsmath,amsthm,amssymb,amsthm,mathrsfs}
\usepackage{feyn}
\usepackage{hyperref}

\newcommand{\dd}{\mathrm{d}}
\newcommand{\bbZ}{\mathbb{Z}}

\DeclareMathOperator{\sign}{sign}

\makeindex

\def\<{\langle}
\def\>{\rangle}
\def\(({\left(}
\def\)){\right)}
\def\[[{\left[}
\def\]]{\right]}

\newcommand{\beq}{\begin{equation}}
\newcommand{\eeq}{\end{equation}}
\newcommand{\beqd}{\begin{displaymath}}
\newcommand{\eeqd}{\end{displaymath}}
\newcommand{\beqa}{\begin{eqnarray}}
\newcommand{\eeqa}{\end{eqnarray}}

\newcommand{\comment}[1]{}

\newcommand{\tB}{\text{\tiny{Bethe}}}

\begin{document}
\title{Loop expansion around the Bethe approximation through the $M$-layer construction}
\author{Ada Altieri $^{1,3}$, Maria Chiara Angelini $^{1}$, Carlo Lucibello $^{2,7}$,  Giorgio Parisi $^{1,4,5}$, Federico Ricci-Tersenghi $^{1,4,5}$ and Tommaso Rizzo $^{1,6}$}

\affiliation{$^1$Dipartimento di Fisica, Sapienza Universit\`a di Roma, Piazzale A. Moro 2, I-00185, Rome, Italy}
\affiliation{$^2$ Politecnico di Torino, Corso Duca degli Abruzzi, 24, I-10129 Torino, Italy}
\affiliation{$^3$ LPTMS, CNRS, Univ. Paris-Sud, Universit\'e Paris-Saclay, 91405 Orsay, France}
\affiliation{$^4$ Nanotec-CNR, UOS Rome, Sapienza Universit\`a di Roma, Piazzale A. Moro 2, I-00185, Rome, Italy}
\affiliation{$^5$ INFN-Sezione di Roma 1, Piazzale A. Moro 2, 00185, Rome}
\affiliation{$^6$ ISC-CNR, UOS Rome, Universit\`a "Sapienza", Piazzale A. Moro 2, I-00185, Rome, Italy}
\affiliation{$^7$ Italian Institute for Genomic Medicine, Via Nizza 52, I-10126 Torino,
Italy}

\begin{abstract}
For every physical model defined on a generic graph or factor graph, the Bethe $M$-layer construction allows building a different model for which the Bethe approximation is exact in the large $M$ limit and it coincides with the original model for $M=1$.
The $1/M$ perturbative series is then expressed by a diagrammatic loop expansion in terms of so-called fat-diagrams.
Our motivation is to study some important second-order phase transitions that do exist on the Bethe lattice but are either qualitatively different or absent in the corresponding fully connected case. In this case the standard approach based on a perturbative expansion around the naive mean field theory (essentially a fully connected model) fails. 

On physical grounds, we expect that when the construction is applied to a lattice in finite dimension there is a small region of the external parameters close to the Bethe critical point where strong deviations from mean-field behavior will be observed. In this region, the $1/M$ expansion for the corrections diverges and
it can be the starting point for determining the correct non-mean-field critical exponents using renormalization group arguments.  
In the end, we will show that the critical series for the generic observable can be expressed as a sum of Feynman diagrams with the same numerical prefactors of field theories. 
However, the contribution of a given diagram is not evaluated associating Gaussian propagators to its lines as in field theories: one has to consider the graph as a portion of the original lattice, replacing the internal lines with appropriate one-dimensional chains, and attaching to the internal points the appropriate number of infinite-size Bethe trees to restore the correct local connectivity of the original model. 

The actual contribution of each (fat-)diagram is the so-called line-connected observable, that includes also contributions from sub-diagrams with appropriate prefactors.
In order to compute the corrections near to the critical point, Feynman diagrams (with their symmetry factors) can be read directly from the appropriate field-theoretical literature; also the computation of momentum integrals is quite similar; the extra works consists in  computing the line-connected observable of the associated fat-diagram in the limit of all lines becoming infinitely long.
\end{abstract}

\maketitle

\section{Introduction}

The Bethe approximation is an essential tool in the context of statistical mechanics as it allows to obtain analytic, albeit usually approximate, results for a huge number of problems.
The Bethe approximation is also quite important in a variety of problems in modern computer science, given its conceptual and practical equivalence to the so-called Belief Propagation algorithm \cite{Yedidia2003}.

The Bethe approximation is exact on tree-like topologies, and it is hindered by the presence of loops.
In this context one is typically interested in improving its predictions, more or less systematically, by taking into account the presence of loops in the lattice associated to the given problem \cite{Yedidia2003,Rizzo2005,Parisi2006,Chertkov2006,Chertkov2006a,Mooij2007,Rizzo2007} or by improving its consistency with linear responses \cite{raymond2017improving} which is usually not satisfied in the presence of loops \cite{wainwright2008graphical}. The main problem is that in many interesting problems the Bethe approximation is non-perturbative, in the sense that there is no small parameter that can be used to develop an expansion to compute corrections \cite{Efetov1990, SackstederIV2013}.

In this work, we discuss the $M$-layer construction. In a nutshell, we start by constructing $M$ copies of the original problem. In the case of a system with quenched random disorder, like random magnetic fields, the disorder is extracted independently and with the same distribution in all the $M$ copies.
Specializing to a two-body problem, for each link of the original problems we have now $M$ copies of the link connecting sites on same layers. In the next step we rewire those links creating inter-layer  connections with some random permutations. We find that in the large $M$-limit
the Bethe approximation, which generally fails on non-tree topologies (as could be the case of $M=1$, that corresponds to the original lattice), becomes asymptotically {\it exact}.
Furthermore, the quantity $1/M$ provides the small parameter to build a perturbative expansion around the Bethe approximation.

Some time after we came up with this construction, we discovered that it was introduced some years ago by Vontobel \cite{Vontobel2013} in the computer science  literature and there is ongoing work in computing the $1/M$ expansion rigorously \cite{Mori2012,Mori2013a}. In this work, we consider the computation of $1/M$ corrections in a systematic way through the so-called cavity method.
We argue that the computation can be carried on in a diagrammatic way introducing "fat-diagrams", topological structures analogous to Feynman diagrams but preserving the local finite-connectivity nature of the original lattice. For instance, while Gaussian propagators are needed to evaluate Feynman diagrams, one-dimensional chains are needed to evaluate fat-diagrams \cite{Parisi2012, RTM14}.

The cavity method, although less rigorous than Vontobel's approach, has the advantage of being more intuitive and efficient for many different types of problems. The main difference with the preceeding works on the $M$-layer construction in the computer science community is that we take an essentially different perspective. 
In fact, we are not particularly interested in evaluating exactly $1/M$ corrections with the goal of solving the actual model ($M=1$) starting from the Bethe solution ($M=\infty$).
We aim instead at using it as a tool to study critical phenomena in finite dimensional systems. In this regard, thanks to universality,  critical exponents depend only on the physical dimension and therefore it is the same to study them in the $M=1$ case or in the large (but finite) $M$ case.
In particular, we will consider large values of $M$ and keep it large, then we will focus on a small region near the Bethe critical point, the critical region, where we expect to see deviations from mean-field behavior. 
 In the critical region all $1/M$ corrections to mean-field behavior diverge and the correct non-MF critical exponents have to be computed through renormalisation.
 
In the modern theory of second-order phase transitions, critical exponents are evaluated starting from continuum field theories rather than microscopic models \cite{Parisi1988,LeBellac1991,Zinn-Justin2002,Itzykson1980,Cardy1996}. This is typically justified invoking universality, but it is also possible to establish a more direct connection between microscopic models and continuum field theories by a closely related construction:  the fully-connected (FC) $M$-layer \cite{Brezin1976,Zinn-Justin2002}.
In the first section of the paper, we will discuss again this classic construction showing how the Landau-Ginzburg Hamiltonian can be derived from the Ising model.   This is a kind of intermediate step because in the actual computation one evaluates the loop expansion of the field theory in order to evaluate the critical exponents.

Compared to fully connected models, that are typically simpler to solve, the Bethe lattice has the advantage of being more similar to realistic finite-dimensional models due to its finite connectivity. On the other hand from the point of view of critical phenomena, a phase transition on the Bethe lattice is mean-field in nature thus, if such a transition does also exist in the large-connectivity/FC limit then the critical exponents should be the same of the corresponding continuum field theory. Our motivation stems from the fact that there are important instances of second-order phase transition that display essential differences when studied on the Bethe lattice rather than on the FC lattice.

To start with, in some cases the transition that we want to study in finite dimensions is just not present in the FC model. 
This is the case for the spin glass (SG) in an external field. In the FC model, there is a transition line in the field-temperature plane, the so-called de Almeida-Thouless line \cite{Parisi1987}. 
This line tends to an infinite value of the field when $T\to 0$; it implies that at $T=0$ there is not a transition in a field and the system is always in the SG phase where replica symmetry is broken. 
This is not the case in finite dimensions, where we know that at $T=0$ and high enough field there is no SG phase. While we lack a proof of the existence of an SG 
transition in a field and the subject is an active research theme \cite{Janus12,Baity-Jesi2014a,Baity-Jesi2014}, it is well-known that things are different from the FC case and more similar to models defined on a Bethe lattice.
Instead, on the Bethe lattice at $T=0$ there is a transition at a finite field $h_c$ (that can be well studied \cite{parisi2014diluted}) and we can think to perform an expansion around this model to understand the fate of this transition in finite dimensions.
More generally, it is known that disordered spin models that in the FC case display some type of SG transition (especially the so-called one-step replica symmetry breaking transition), 
have often a quite different behaviour in finite dimensional systems, often more similar to the Bethe version \cite{Cammarota2013}.

Another example is Anderson localization. The existence of a transition is prohibited by definition in the FC case: if each site is linked to all the others, it can not exist a localized state. Things are different in Bethe lattice where, thanks to the finite connectivity,
a localization transition does exist  \cite{Abou-Chacra1973,Abou-Chacra1974,Biroli2010}. In this case, the behavior of the transition is not known in finite dimensions, and it is unknown the upper critical dimension. One would like to study the finite dimensional problem starting from the mean-field solution and developing a loop expansion around the Bethe approximation. 
This has motivated earlier efforts by Efetov \cite{Efetov1990} to develop a topological (loop) expansion in terms of fat-diagrams around the Bethe approximation, see also \cite{Parisi2006}. 

In some other cases, the transition in the FC model is present but it appears to have significant differences with respect to the one in finite dimension. 
This is the case for the Random Field Ising model (RFIM). In this case, the FC MF solution and the corresponding loop expansion is well understood \cite{DeDominicis2006}. This expansion implies dimensional reduction meaning that the critical exponents of RFIM in $D$ dimensions are the same of a pure ferromagnet in $d=D-2$ dimensions \cite{Sourlas1979}.
However, it is well known that for low enough dimensions dimensional reduction is not valid: the breaking is due to non-perturbative effects. 
For the RFIM, we can write a self-consistent equation
for the local magnetization that has a unique solution only in the FC case. The assumption of a unique solution is what leads to dimensional reduction.
However, we know that at finite connectivity multiple solutions exist. This is a property in common between finite dimensional lattices and Bethe lattices. 
One possibility is that the loop expansion around the FC fixed point leads to wrong results precisely because it does not account correctly for the presence of multiple solutions and therefore
expanding around the Bethe solution, that displays a fixed point with multiple solutions, could give different results \cite{Sourlas1979}.
 
More generally we can think to perform an expansion around the Bethe solution for whatever problem can be studied on the Bethe lattice. This includes, of course, statistical mechanics models (with or without quenched disorder) at finite temperature but also systems where there is no Hamiltonian at all like percolation, $k$-core percolation or zero temperature systems. 

The lack of a reliable FC model for the various second-order phase transition mentioned above reflects naturally in the lack of reliable continuum field theory 
to be eventually studied by a loop expansion. 
As we will see in the following, the Bethe $M$-layer construction allows obtaining the loop expansion (and then the critical exponents) directly without knowing what is the underlying continuum field theory of the problem. For instance, we will show how to recover the loop expansion for percolation, without resorting to the analytic continuation of the Potts model. 

The paper is organized as follows. In sec (\ref{FCMlayer}) we discuss the FC $M$-layer construction.
We will be working with very large values of $M$ and although the resulting model seems somehow academic and looks rather different from the original $M=1$ model, universality grants that their critical behavior is the same.
We will show explicitly that in a region close to the MF critical point the system for large $M$ is described exactly by a continuum field theory.
In particular, we will recall that all corrections in powers of $1/M$ are divergent at the critical point and that the leading divergences at each order are exactly given by the corresponding terms in the loop expansion of the continuum field theory.

In section (\ref{sec:M-layer-Bethe}) we will introduce the Bethe $M$-layer construction arguing that in the large $M$ limit the Bethe solution becomes exact.
We will then show how to compute the leading order correction to the two-point correlation by writing it as a sum over non-backtracking walks on the original lattice and we will focus on its divergent behavior close to the critical point.

In section (\ref{cavity-M-layer}) we discuss the computation of $1/M$ corrections for a generic observable in a systematic way introducing the notion of fat-diagrams, that are essentially subgraphs of the original lattice.
Firstly, we introduce a general graphical expansion in terms of fat-diagrams and then we show that within the $M$-layer construction the expansion is a loop expansion in the sense that the contribution of each fat-diagrams is proportional to $1/M^L$ where $L$ is the number of topological loops in the diagram.
We show that the contribution of each fat-diagrams is not trivially given by the value of the given observable on the corresponding graph but requires the subtraction and addition of other terms with appropriate coefficients. The actual contribution is indeed given by the so-called line-connected observables.

We then turn to study the critical behavior deriving Feynman rules for the expansion. The problem is then reduced to the evaluation of line-connected observables on fat-diagram close to the critical point and when all distances are large.
In the last part of the section, we give an explicit expression for line-connected observables and we use it to prove that, 
whenever the Bethe phase transition {\it is} the same of the FC system, 
the loop expansion at criticality is exactly the same of the corresponding continuum field theory. In particular, we consider concrete examples that match to ordinary $\phi^3$, $\phi^4$ theories.
This is a kind of consistency check of the whole approach prior to non-trivial applications (to be presented elsewhere) to problems where the phase transition on the FC lattice is either different or absent.
In the last subsection, we will discuss some technical issue associated with the fact that,  at variance with the FC $M$-layer construction, the Bethe $M$-layer construction introduces some spurious disorder in the problem. In particular, we will address the issue of this spurious disorder on critical behavior.
Finally in Sec. \ref{Conclus} we conclude providing a general review and underlying the key points of the paper.

\section{The Fully Connected $M$-layer construction}
\label{FCMlayer}

In this section we will recall some basic properties of the field-theoretical loop expansion.
Many of the properties we want to underline can already be discussed at the simplest level, {\it i.e.} in zero dimension.
Let us consider the paradigmatic ferromagnetic transition in the fully-connected Curie-Weiss model:
\begin{equation}
\mathcal{H}(\sigma)=-\frac{J}{2N}\sum_{i,j=1}^N \sigma_i \sigma_j -h \sum_{i=1}^N \sigma_i \ .
\end{equation}
By standard manipulation the partition function of the system of $N$ Ising spins can be exactly written as an integral of an action:
\begin{equation}
Z = \left( \frac{\beta J N}{2 \pi }\right)^{1/2} \int_{-\infty}^{+\infty} {dm\ e^{ -N \beta f_\beta(m, h)}  },
\label{eq:part_funct_ising}
\end{equation}
where the action reads:
\begin{equation}
f_\beta(m, h)=\frac{1}{2} J m^2 -\frac{1}{\beta} \log \left[ 2\cosh(\beta J m+ \beta h) \right] \ .
\label{eq:free_energy_ising}
\end{equation}

The mean-field (MF) approximation, which is exact in the Curie-Weiss model for large $N$, amounts to approximate the above integral with the value at the global minimum of the action. 
Let us consider the $h=0$ case. Corrections to MF  can be computed systematically writing the action as
\begin{equation}
Z \propto  \int_{-\infty}^{+\infty} {dm \cdot e^{ -N \left(\tau \,m^2+g_4\,m^4+g_6\,m^6+... \right)}  }
\label{eq:part_funct_1/N}
\end{equation}
where we expanded the last term of eq. (\ref{eq:free_energy_ising}). The reduced temperature is defined as $\tau=\frac{\beta J}{2} \left( 1-\beta J\right)$ ( the critical temperature corresponds to $\tau=0$) 
and the coupling constants are $g_4=\frac{\beta^4J^4}{12}$ , $g_6=-\frac{\beta^6J^6}{45}$, $\dots$.
This generates the loop expansion where each term is associated with a Feynman diagram. To each Feynman diagram, there is a factor $1/N$ for each line and a factor $N$ for each vertex leading to $1/N$ to the power $L$, where $L=I-V+1$ is the number of loops, $I$ being the number of lines and $V$ the number of vertexes.

When $N$ is $O(1)$, that is for small system sizes, the loop expansion is not accurate because corrections are of the same order of the leading (saddle point) order. If we consider though  large values of $N$ at fixed value of the temperature, the MF approximation becomes increasingly accurate and the loop expansion is a perturbative expansion where terms with higher number of loops give smaller corrections.
Note that, at a given order in the loop expansion, all types of vertexes contribute to the expansion, and not just those with lower degrees. Therefore the full $1/N$ series depends on the infinite set of coupling constants $\{g_4, g_6, g_8, \dots \}$.

There is a problem however with the above result: the MF approximation predicts a phase transition at $\tau=0$ but at finite $N$ there cannot be any phase transition: the partition function is analytic in any of the external parameters. 
Instead, the transition is present at all orders in the $1/N$ expansion, implying that its disappearance is a non-perturbative phenomenon.
In fact one can see that the corrections in powers of $1/N$ are all divergent at the critical point $\tau=0$,  because the propagator is given by $1/(N \tau)$. 
Therefore, at fixed number of loops $L$, all diagrams are $O(1/N^L)$ but the divergence is $O(1/\tau^I)$. This means that the most diverging diagrams are those with the highest number of lines. 
One can easily see that the number of lines is maximized at fixed $L$ by the diagrams with vertexes with the lowest possible degree (four in this case). 
This implies that, {\it while the $1/N$ corrections depend on the whole infinite set of coupling constants $\{g_4, g_6, g_8, \dots \}$, the leading divergences are only controlled by the $g_4$ coupling constant}.

For these vertexes we have $I= 2V \rightarrow L=I/2+1$. Defining $\tau=\tilde{\tau} N^{-1/2}$, we see that the loop expansion becomes an expansion in powers of $1/\tilde{\tau}$. In the end, we want to evaluate this series at $\tau=0$ by means of a re-summation procedure.
The condition $ \tilde{\tau} =O(1) $ or equivalently $\tau=O( N^{-1/2})$ defines the critical region where the MF behavior does not hold and it is the zero-dimensional equivalent of the Ginzburg criterion. Outside this region the MF approximation is accurate, while in this region it is not, in agreement with the fact that, 
if we go to $\tau=0$ at fixed albeit large $N$, we must see that there is no phase transition.

In the critical region the loop expansion is non-perturbative, but on the other hand, we expect that the problem is somehow simpler than the original one, since the leading divergences depend solely on the $g_4$ coupling constant.
Indeed by performing  the simple rescalings $N^{1/4}m \rightarrow m$, $\sqrt{N}\tau \rightarrow \tau$, we can eliminate the $N$ dependence from the first terms:
\begin{equation}
Z \propto  \int_{-\infty}^{+\infty} {dm \cdot e^{ -\tau m^2-g_4m^4-g_6\frac{1}{\sqrt{N}} \, m^6+... }  }.
\end{equation}
We see that the {\it in the critical region we must not evaluate the whole integral, but we have to retain only the quadratic and quartic term because the others give subleading corrections}. 

We make two more considerations before moving to the fully connected $M$-layer construction for finite-dimensional models.
Firstly, we note that considering a different model, {\it e.g.} a soft-spin model, would change all the coupling constants and the whole series would be different outside the critical region. Nevertheless, all models are described by the {\it same} critical theory in the critical region once the rescaled variables $m$ and $\tau$ are further rescaled to eliminate the $g_4$ dependence. This is the zero-dimensional analog of universality as it occurs for a genuine phase transition in finite dimension. 
Secondly, we note that for the ferromagnetic transition the zero-dimensional theory $\int dm\ e^{-\tau m^2- g m^4}$ can be studied directly. However in more complex situations, {\it e.g.} disordered systems, the zero-dimensional integral cannot be studied directly and one must resort again to a loop expansion. In the end, one faces the problem of re-summing the series in powers of $1/\tau$ in order to get accurate results at small values of $\tau$.

The fully connected $M$-layer construction in finite dimension has many similarities with the zero-dimensional problem. Let us consider a standard Ising model on a finite dimensional lattice:
\begin{equation}
H=-\frac{1}{2}\sum_{i,j=1}^{N} J_{ij}\, \sigma_i\,\sigma_j - \sum_{i=1}^{N} h_i \sigma_i,
\end{equation}
where the matrix $J$ also defines the structure of the lattice (i.e. $J_{ij}=0$ if vertexes $i$ and $j$ are not neighbours).
On each site of the lattice, we put a stack of $M$ Ising spins. At this point each spin is coupled ferromagnetically with the $M$ spins on each of its $2D$ nearest neighbors. 
We rescale the couplings by a factor $M$, to have a good $M\rightarrow \infty$ limit. We denote with $\{\sigma_i^\alpha\}$ the configurations of the augmented system, 
and write its partition function as
\begin{equation}
Z_M = \sum_{\{\sigma^\alpha_i\}} \exp\left\{\frac{\beta}{2M}\sum_{i,j}^N J_{ij}\sum_{\alpha,\alpha'}^M \sigma^\alpha_i\sigma^{\alpha'}_j+\beta\sum_{i}^N h_{i}\sum_{\alpha}^M \sigma^\alpha_i\right\},
\end{equation}
which with few manipulation can be rewritten as
\begin{equation}
Z_M \propto \int_{-\infty}^{+\infty} \prod_{i=1}^{N} \text{d} m_i\  \exp{  \left \lbrace M \left[ -\frac{\beta}{2} \sum_{ij} m_i (J^{-1})_{ij} m_j +\sum_i \log {\left[ 2 \cosh(\beta m_i +\beta h_i) \right] } \right] \right \rbrace }.
\label{Zm-fc}
\end{equation}
For $M=1$ we recover the original system. Eq. \eqref{Zm-fc}  for $M=1$ appeared for the first time in the literature in Ref. \cite{Polyakov1969} an for the first time (see also Ref. \cite{Langer1967}). 

The mean field approximation of statistical physics can be interpreted as the limit of large $M$ of the fully-connected $M$-layer construction, which amounts to saddle point evaluation of the above integral.
The propagator for the augmented system is $O(1/M)$ while the vertexes contribute with a $O(M)$ factor. A Feynman diagram with $L$ loops has a prefactor proportional to $1/M^L$. 
This implies that in the original system ($M=1$) the MF approximation is not accurate. This is well-known: for instance, the actual critical temperature of the 3D Ising model is quite different from its MF value.
Instead, if we consider large values of $M$ at fixed temperature the MF approximation and the corresponding $1/M$ expansion becomes accurate.

Each term in the $1/M$ expansion diverges close to the MF critical temperature. Once again this is in agreement with the expectation that at fixed albeit large $M$  deviations from MF behavior must occur in the critical region. This must happen either because the MF transition is washed out, as in dimensions  $d=1$ and $d=0$, or because the critical exponents must be different from the MF ones, {\it i.e.} for $d=2$ and $d=3$.

Let us consider for simplicity the zero magnetic field case. 
Similarly to what we saw in zero dimensions, after expanding the last term in Eq. (\ref{Zm-fc}),
inspection of the diagrams reveals that, while at fixed temperature the $1/M$ expansion depends on all coupling constants $\{ g_4, g_6, g_8 , \dots \}$ and on the exact expression of the free propagator (in momentum space)
\begin{equation}
\hat{G}_0(k)=\frac{1}{1-\frac{1}{d} \sum_{\mu=1}^{d} \cos(k_\mu) } =\left( 1-2 d \beta +\beta \sum_{\mu=1}^{d} k_\mu^2+ O(k_\mu^4) \right)^{-1} \ ,
\end{equation}
the leading divergences are obtained considering only the quartic vertexes and the short-distance behavior of the propagatotor, i.e. neglecting $k^4$ and higher order terms  in its expression.
In other words {\it in the critical region near the MF critical temperature the relevant divergences are the same that would be obtained by the loop expansion of the Ginzburg-Landau theory}. 
As we did before, this can be also seen directly starting from the expression of the action (in momentum space)
\begin{equation}
Z_M \propto \int_{-\infty}^{+\infty} \prod_{|k| < \Lambda} d \tilde{m}(k) d \tilde{m}^*(k)\ \exp{  \left \lbrace M \left[  -\int d^d k\ 
(\tau + c_2 k^2+ c_4 k^4+\dots ) |\tilde{m}(k)|^2- g_4 \sum_i m_i^4 + \dots \right] \right \rbrace },
\end{equation}
where $\tilde{m}(k)$ is the Fourier transform of $\{m_i\}$ and  $\Lambda$ is the momentum cut-off, {\it i.e.} the inverse of the lattice spacing.
For $d<4$ we perform the following rescalings:
\beq
x \, M^{-{1 \over \epsilon}} \rightarrow x ,\  \tau \, M^{{2 \over \epsilon}} \rightarrow \tau ,\  m \, M^{{1 \over \epsilon}} \rightarrow m \, , 
\eeq
where $\epsilon \equiv 4-d$.  We obtain an expression that does not depend on $M$, plus small corrections:
\beqa
Z_M & \propto & \int_{-\infty}^{+\infty} \prod_{k} d \tilde{m}(k) d \tilde{m}^*(k)\  \exp  \left [  - \int d^d k\ 
(\tau + c_2 k^2) |\tilde{m}(k)|^2 - g_4 \int d^d x \,m(x)^4 + \right.
\nonumber
\\
&  &-  \left.  \sum_{s>4} g_s \  O \, \left(M^{-\frac{s-4}{\epsilon}} \right)  - \sum_{s >2} c_s \ O \, \left(M^{-\frac{s-2}{ \epsilon}} \right) \right]
\eeqa
where we have used $k \, M^{{1 \over \epsilon}} \rightarrow k $ and $\tilde{m} M^{-{d-1 \over \epsilon}} \rightarrow \tilde{m}$. Note that the rescaled cut-off has gone to infinity, meaning that we have obtained the continuum limit of the Ginzburg-Landau theory.

Summarizing, the fully connected $M$-layer construction has the following properties:
\begin{itemize}

\item in the large $M$ limit at {\it fixed} temperature the MF approximation is accurate and the loop expansion is perturbative. In this region, all coupling constants are relevant for the $1/M$ expansion.

\item Deviations from MF behavior are observed in a region centered at the MF critical temperature that shrinks with $M$ as $1/M^{2/ \epsilon }$. Correlation functions are Gaussian at short distances also in the critical region while non-trivial behavior is observed at large distances. More precisely any $n$-points correlation function at large distances, $O(M^{1/\epsilon})$, and small deviations  from the MF critical temperature, $O(1/M^{2/ \epsilon })$, behaves as $1/M^{n/\epsilon}$ with a scaling function that is determined by the Ginzburg-Landau theory in the continuum limit.
Therefore only the quartic coupling constant is relevant.

\end{itemize}

The study of the Ginzburg-Landau Hamiltonian in the continuum limit leads to the well-known problem of renormalization. 
We know that the continuum limit is not well defined for $d \geq 4$. On the other hand, the above scalings are meaningless for $d > 4$, while in $d=4$ they just tell us that the size of the region of deviations from MF behavior vanishes (since  $\epsilon=0$).

The continuum limit instead is well-defined in $d<4$. However, if we try to compute correlations by a loop expansion
we discover that all coefficients of the series are divergent also at large values of $\tau$ due to the infinite cut-off implicit in the continuum limit. The renormalization procedure instead tells us how to reshuffle the result in order to obtain finite results \cite{Parisi1988}.

In the following, we will consider the Bethe $M$-layer construction. In this case, the large $M$ limit of the theory will be given by the Bethe solution and we will discuss how to compute $1/M$ corrections.
With critical phenomena in mind, we will study the divergence of the corrections at each order as we approach the Bethe critical temperature, with the goal of using these series to extract the non-MF critical exponents in finite dimension.

\begin{figure}[h]
\includegraphics[width=1\textwidth]{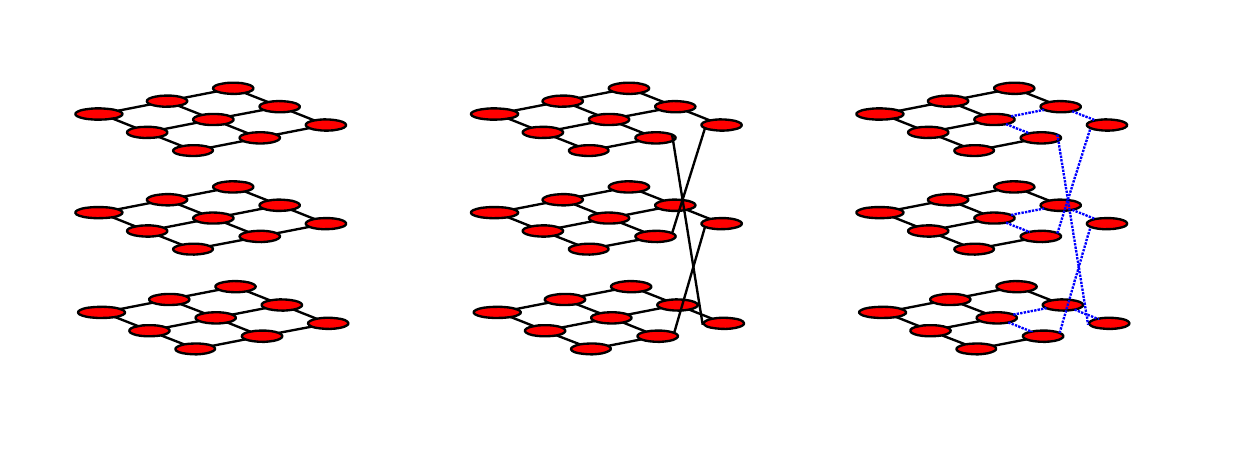}
\caption{Qualitative representation of a $M$-layer construction for a 2D regular lattice with $M=3$. 
(left) the original lattice replicated $M$ times;
(center) for a specific edge in the original graph, a possible rewiring of its $M$ copies;
(right) a simple loop in the $M$-graph (dashed lines), corresponding to a non-backtracking closed path, once projected onto the original lattice.}
\label{Mlayers}
\end{figure}

\section{The Bethe M-Layer}
\label{sec:M-layer-Bethe}
We now introduce the Bethe M-layer construction, the main subject of the present paper. In order to preserve an high degree of generality and to connect our approach to Vontobel's graph cover formalism \cite{Vontobel2013}, we shall consider the topological structure of a generic factor graph \cite{Montanari2009}. 

A factor graph $g$ is a bipartite graph with two nodes sets, $V$ and $F$, whose elements are called factor nodes and variable nodes respectively. The edge set $E$ of $g$ has elements of the for $(a,i)$ with $a\in F$ and $i\in V$.  A factor graph can also be interpreted as an hypergraph, identifying factor node with hyperedges. If each factor node has degree 2, the factor graph is equivalent to a graph whose edges $(i,j)$ correspond to factor nodes. The adjacency matrix of $g$ is 
\begin{equation}
C_{ai}=\begin{cases}
1 & (a,i)\in \text{if } E \\
0 & \text{otherwise}
\end{cases}
\end{equation}

Given a factor graph $g=(V,F,E)$, the Bethe M-layer construction on $g$ is a new factor graph $g'$ built in the following two steps:
\begin{itemize}
\item create a factor graph $g'$ made of $M$ copies of the variable and factor nodes of $g$, labeled with $(i,\alpha)$ and with  $(a,\alpha)$ respectively,  where $i\in V, a\in F,\alpha\in [M]$;
\item for each of the original edges $(a,i)$ choose uniformly and independently at random a permutation $\pi_{ai}$ of the set $[M]$, i.e. $\pi_{ai} : [M]\to [M]$. Let us call $\Pi$ the set of all such permutations, $\Pi =\{\pi_{ai} :(a,i)\in E\}$. The adjacency matrix $C^\Pi$ of $g'$ is then given  by $C^\Pi_{a\alpha,i\beta}=C_{ai}\,\delta_{\beta,\pi_{ai}(\alpha)}$.
\end{itemize}
Here we used the notation  $[M]\equiv\{1,\dots,M\}$. In the particular case of standard graphs, the above procedure consists, for each edge $(i,j)$, in choosing one of the $M!$ 
 possible matchings among the $M$ copies of $i$ and $j$, as depicted in Fig. \ref{Mlayers}.
 
A model average over all the realization of the permutations $\Pi$ (\textit{rewirings}) has to be considered in the end. We will denote such average as $\langle \bullet \rangle_{rew}$, and discuss its peculiarities later in this paragraph and in Section \ref{sec:spurious}.

Given that the random permutations of the layer indexes are independent for each link $(a,i)$ of the original graph, one can be easily convinced that in the large $M$ limit the resulting graph has a locally tree-like structure (the density of loops is proportional to $1/M$) 
and therefore the Bethe approximation must be increasingly accurate for increasing values of $M$.
The effect of a finite $M$ is to introduce loops in the model, inducing corrections on the Bethe result. One of such loops is shown in the right part of Fig. \ref{Mlayers}.

The main difference of this procedure with respect to the FC $M$-layer discussed above  is that the 
connectivity of the model remains the original one, e.g.  $2D$ on a hypercubic $D$-dimensional lattice.
Note that up to this point we have not specified the Hamiltonian of the system and indeed another very interesting feature of the Bethe $M$-layer construction is that 
it does not require a Hamiltonian. This implies that one can use it also in a context where there is no Hamiltonian at all, {\it e.g.} percolation or purely dynamical models like the Kob-Andersen model \cite{Boccagna2017} or the Kardar-Parisi-Zhang surface growth model \cite{Kardar1986}.

For Hamiltonian systems, some additional interesting points can be discussed.
In order to be definite let us consider the case of a pairwise  Ising model where neighboring spins interact through a coupling $J_{ij}$. Later we will again discuss the problem in full generality. In this case for each realization of the $M$-lattice, we have a free energy and we can either consider white averages $\langle \ln \,Z \rangle_{rew}$ over all possible rewirings, or annealed averages, $\langle Z \rangle_{rew}$. The latter amounts to weight each realization of the $M$-layer with its free energy. One can choose to consider annealed or quenched averages over the rewiring depending on what is more convenient. Vontobel's approach involves an annealed average \cite{Vontobel2013}.

In the context of systems that are natively disordered, e.g. spin glasses (SG), another useful trick is to first generate an instance of the $M$-layer lattice and then assign the  $J_{ij}^{\alpha\beta}$ as independent random variables. 
Within the replica formalism often used to deals with such systems \cite{Parisi1987},
this corresponds to first introduce replicas in the original system, average over disorder,  and then apply the $M$-layer construction. If these steps are inverted all $J_{ij}^{\alpha\beta}$ corresponding to the same link would be the same, and the analysis would be much more involved.

For the Ising model in finite dimensions, analytical progress can be made by means of techniques similar to those used for random regular graphs (RRG) \cite{RRG14}. 
In particular one can derive the (quenched or average) free energy as an integral of an action that has a factor $M$ in front of it. 
Much as in the FC construction, this justifies the use of a saddle-point approximation with the difference that, in this case, 
the saddle point free energy is the  Bethe one, which is essentially Vontobel's result \cite{Lucibello2015, Vontobel2013}.
 
Then one could develop an $1/M$ expansion by systematically computing the corrections around the saddle-point approximation. While this programme is conceptually clear, the first steps being taken in Ref. \cite{Lucibello2015}, it is also plagued by technical difficulties, one of such being the presence of null modes \cite{Lucibello2015}. This a scenario which generally arises in bipartite matching models \cite{Mezard1987, Ratieville2002} and also in other expansions around the Bethe solution \cite{Parisi2006,Chertkov2006}.

In this respect, the cavity method \cite{Parisi1987} has many advantages. In particular,  the meaning of the various corrections is more transparent.  Most importantly it can be applied also when there is no Hamiltonian and thus no saddle-point computation.
In the following, we will show how to find a well defined $1/M$ expansion using the cavity method, and we will see
that it corresponds to a loop expansion, where this time the loops are spatial loops.
We will start discussing the leading correction for the correlation in a soft-spin ferromagnetic model in the following subsection. This will serve as an illustration of the more general treatment that will be given later.

\subsection{Example: The leading order $1/M$ corrections to two-point correlations}
 \label{sec:leadingorder}

To be definite the original ($M=1$) model is a pairwise soft-spin model on the regular lattice in finite dimension. We denote with $x$ and $y$ two distinguished vertexes on the lattice.
We will discuss the leading $1/M$ correction to the connected correlation function between two spins $\sigma_x^\alpha$ and $\sigma_y^\beta$,  where the Greek indexes label the $M$ layers.
Later we will see how to generalize the discussion to all possible observables and all orders.

This correlation is a random quantity that depends on the realization of the $M$-lattice. 
The key point of the whole analysis is that in the large $M$ limit a given realization of the $M$-lattice looks tree-like locally with high probability.
Thus for any $x$ and $y$ the two spins $\sigma_x^\alpha$ and $\sigma_y^\beta$ are not correlated in the $M=\infty$ limit once we have rewired the links: the correlation is simply zero at leading order.

Small corrections to the average value at finite $M$ are due to the rare re-wirings in which the two spins happen to be close to each other.
The simplest such possibility occurs when the spins are connected by a simple path of finite length $L$. A simple path is a sequence of adjacent vertexes in the $M$-graph such that none of the vertexes is repeated in the sequence. In what follows we will drop "simple" and just call them paths.
 
At leading order the probability that a given path of length $L$ is present is given by $M^{-L}$, since every link of the path is present with probability $1/M$. On the other hand, there are many such paths. In order to count them, it is convenient to order them according to their projection on the original lattice. 
Given a path on the original lattice between the points $x$ and $y$, the total number of paths connecting $\sigma_x^\alpha$ and $\sigma_y^\beta$ 
in the corresponding $M$ layer model is $M^{L-1}$ at leading order, since each of the $L-1$ internal vertexes can be chosen in $M$ different layers.
Therefore, the total weight of the paths that have the same projection on the original lattice is $1/M$.
It is also evident that the correlation between the two spins can be approximated by the corresponding correlation on a Bethe lattice between two spins at distance $L$. 

It is thus natural to order the paths according to their length and to sum over all paths of length $L$ connecting point $x$ and point $y$ on the original lattice. 
Notice that the projection of a simple path in the $M$-lattice onto the original lattice is no more simple: vertex repetitions are allowed, the only condition that any three consecutive vertexes have to be pairwise different. This corresponds to the definition of non-backtracking path \cite{Fitzner2013}.

Therefore, one should sum over the non-backtracking paths on the original lattice because,  given one site $(x,\alpha)$ of the layered graph, there is only one value of 
$\beta=1,\dots, M$ such that the site $(z,\beta)$ - with $z$ neighbor of $x$ in the original lattice - is connected to $(x,\alpha)$.
Let $P(\sigma_x^\alpha,\sigma_y^\beta)$ be the marginal distribution over spins $\sigma_x^\alpha$ and $\sigma_y^\beta$ for a given $M$ graph realization, obtained by tracing the Boltzmann distribution over all the other spins. Marginal distributions over any other subset of variables have  similar definitions.
We define the
connected distribution as:
\beq
P_c(\sigma_x^\alpha,\sigma_y^\beta) \equiv P(\sigma_x^\alpha,\sigma_y^\beta)-P(\sigma_x^\alpha)P(\sigma_y^\beta)\;,
\eeq
Taking the average over the rewirings, it is easy to show that 
\beq
\langle P_c(\sigma_x^\alpha,\sigma_y^\beta) \rangle_{rew}={1 \over M}\sum_{L=1}^{\infty} b_L(x,y) \, P_{c,L}^{\tB}(\sigma_x^\alpha,\sigma_y^\beta)+O\left( {1 \over M^2}\right)\;,
\label{eq:1/M}
\eeq
 where $P_{c,L}^{\text{\tiny{Bethe}}}$ is the connected distribution between two sites at distance $L$ on a Bethe lattice
 and $b_L(x,y)$ is the total number of non-backtracking paths of length $L$ connecting point $x$ and point $y$ on the original lattice.
 For more details about the computation of $b_L$ see Appendix \ref{app:nbw}.
Practically with Eq.~(\ref{eq:1/M}) we are saying that at order $O(1/M)$ in our new expansion,
 the correlation between two points is just the sum of the correlation along all the possible non-backtracking paths linking them (spins are considered as independent if the links on the path are cut).

On the Bethe lattice, correlations are associated with one-dimensional chains. This is because essentially there is only one path connecting them;  if the system is homogeneous in space further progress can be made. To this case one can often reconduct also systems with i.i.d. quenched disorder: using the replica trick one can average over the disorder the replicated partition function $Z^n$, and the resulting system will be homogeneous. The price paid is that of working with $n$-components variables, with $n$ that will be sent to zero at the end of the computation.

Generalizing the computation on one dimensional chains, on a homogeneous Bethe lattice correlations can be computed by transfer matrix methods.  The derivation is presented in Appendix \ref{app:bethe}, here we do a brief recap.
The two point distribution on the Bethe lattice, $P^{\tB}_L(\sigma,\tau)$, can be expressed in term of  functions $a_\lambda(\sigma)$ related to the eigenvectors of a transfer matrix, with eigenvalues $|\lambda |<1$. The   $\lambda=1$ eigenvalue instead is the one associated to the single spin marginal distribution in the Bethe lattice,  $P^{\tB}(\sigma)$, which can be computed also within the Belief-Propagation algorithm. 
The functions $a_\lambda(\sigma)$  satisfy the following property: $\int d\mu(\sigma)\ a_\lambda(\sigma)=0$ for $\lambda \neq 1$.  On the Bethe lattice the connected two point distribution takes the form

\begin{equation}
\begin{aligned}
P^{\tB}_{c,L}(\sigma,\tau)&= P^{\tB}_{L}(\sigma,\tau) - P^{\tB}(\sigma)P^{\tB}(\tau) =\sum_{|\lambda|< 1}\lambda^L \,a_\lambda(\sigma) a_\lambda(\tau).
\end{aligned}
\end{equation}
Last expression is particularly handy when used in conjunction with the generating function of the non-backtracking paths (see Appendix \ref{app:nbw}):
\begin{equation}
 B_\lambda(x, y)=\sum^\infty_{L=1} b_L(x,y)\, \lambda^L,
\end{equation}
so that we can rewrite Eq. \eqref{eq:1/M} as
\beq
\langle P_c(\sigma_x^\alpha,\sigma_y^\beta) \rangle_{rew}={1 \over M}
\sum_{\lambda\neq1} B_\lambda(x,y)\, a_\lambda(\sigma_x^\alpha) a_\lambda(\sigma_y^\beta)+O\left( {1 \over M^2}\right).
\label{eq:Pc1}
\eeq
The exact computation of the $1/M$ corrections requires thus both a precise knowledge of the formula for non-backtracking paths and a precise knowledge of an enventually infinite set of eigenvalues and functions $a_\lambda(\sigma)$. Note that for the Ising model without disorder we would have only one eigenvalue besides $\lambda=1$. In the presence of disorder, one can resort to the replica formalism of \cite{RTM14} to compute the eigenvalues and the associated eigenfunctions.

\subsubsection{Critical Behavior of the two-point correlation at leading order}

Let us now investigate the consequences of the discussion of previous section on critical phenomena.

Approaching a second-order phase transition on the Bethe lattice with connectivity $2D$ the second largest eigenvalue $\lambda'$ (the first being  equal to one) goes to the value 
$\lambda_c \equiv 1/(2D-1)$, that is the inverse of the branching ratio. This corresponds in fact to the divergence of a susceptibility that can be expressed as $\chi \approx \sum_{L=1}^{+\infty} (2D-1)^L \lambda'^L$.

In the translationally invariant original lattice, consider  the fourier transform of Eq. \ref{eq:Pc1}, where both left hand side and right hand side are considered as a function of $x-y$. We obtain: 
\beq
\langle P_c(\sigma,\tau; k) \rangle_{rew}={1 \over M}
\sum_{\lambda\neq1} \hat B_\lambda(k)\, a_\lambda(\sigma) a_\lambda(\tau)+O\left( {1 \over M^2}\right)\,,
\eeq
where $\hat{B}_\lambda(k)$ is the Fourier transform of  $B_\lambda(x-y)$. For small $k$ we have (Appendix \ref{app:nbw})
\begin{equation}
\hat{B}_\lambda(k)  \propto \frac{1}{k^2+m^2}, \qquad m^2=1-(2D-1)\lambda\,.
\end{equation}

Since we are interested in critical behavior, we want to study the large distance (small $k$) behavior  in the region of external parameters where the second largest eigenvalue $\lambda'$ is close to $\lambda_c\equiv1/(2D-1)) $.
On the other hand, the behavior for small $k$ of the Fourier transform of the generating function 
of non-backtracking paths is precisely given by $\hat B_{\lambda}(k)= (\lambda-\lambda_c+k^2+O(k^4))^{-1}$.
This implies that {\it close to the critical point only the second-largest eigenvalue need to be considered}:
\beq
\langle P_c(\sigma,\tau; k) \rangle_{rew}={1 \over M}\sum_{\lambda\neq1} \hat  B_\lambda(k)\, a_{\lambda}(\sigma)a_{\lambda}(\tau)\approx {1 \over M} \hat B_{\lambda'}(k)  a_{\lambda'}(\sigma) a_{\lambda'}(\tau) \ .
\eeq
leading to:
\beq
\langle P_c(\sigma,\tau; k) \rangle_{rew} \approx {1 \over M}  {1 \over m^2+k^2}\, a_{\lambda'}(\sigma)a_{\lambda'}(\tau) \ ,
\eeq
where $m^2 \propto \lambda'-\lambda_c$ (we are using a field theoretical notation) is a linear function of the external parameters vanishing at the critical point.
Thus we see that near the critical point the correlation at leading order in the $M$-layer Bethe expansion has  the same (Gaussian) form that appears naturally in the expansion around the fully connected model.

This is the simplest manifestation of an important feature connected to universality.  In this paper, we take the perspective that the $M$-layer construction is a tool to study critical phenomena by developing a loop expansion. Now if the very same problem can already be studied by the FC $M$-layer construction ({\it i.e.} by a field theory) there is just no need to introduce the Bethe $M$-layer construction.  At most, in this case, the game is to show that the critical Bethe $M$-layer loop expansion is identical to the field theoretical loop expansion. Non-trivial results can only be obtained considering models where the FC construction does not work or there is no evident field theory. 

In the following, we will discuss the critical $1/M$ expansion in full generality and show consistently that the FC results can be recovered at all orders when they are available. As we shall see later, everything is quite simple if only one eigenvalue becomes critical. In the case where an infinite number of eigenvalues become critical, the situation is much more complex and a  more refined computation is needed.

\section{ The Bethe $M$-Layer Loop expansion}
\label{cavity-M-layer}

\subsection{The Graph-Theoretical Loop Expansion} 
\label{SubSec:graph expansion}

In this subsection, we introduce a graph-theoretical expansion in order to write observables on lattices as the sum of contributions corresponding to subgraphs.
The resulting expressions hold in general, for every model on every graph.

However, for a generic graph, the expansion is non-perturbative in the sense that all terms in the expansion need to be evaluated because they have the same order of magnitude. Thus it is of limited use, but as we will see in the following section, 
it becomes a perturbative expansion in the $M$-layer model.

We consider a generic physical process defined on a lattice with two-body interactions.
We will mostly work without specifying the actual nature of the process; in practice, we assume that it is possible to study it on the Bethe lattice.
This includes, of course, statistical mechanics models (with or without quenched disorder) at finite temperature but also systems where there is no Hamiltonian like percolation, $k$-core percolation or zero temperature systems.

Our goal is to compute observables as determined by the physical process. 
Keeping the discussion at a very general level, observables are functions of the positions of the sites on the lattice. Thus a two-point observable is a function 
$A(x,y)$ where $x$ and $y$ are positions and an observable of order $n$ is a function $A(x_1,\dots, x_n)$.
In the high temperature phase (and we approach the critical point from this phase) the clustering property holds, implying that the two-point observable $A(x,y)$ factorizes onto the product of two one-point observables $A(x,y) \approx A(x) A(y)$ when the positions $x$ and $y$ are distant on the lattice: we will thus consider connected observables $A_c(x,y) = A(x,y)-A(x)A(y)$ that vanish when one of its arguments become very large.
Connected observables of generic order $n$, $A_c(x_1,\dots,x_n)$, vanish when at least one of the lattice points is far away from the others, and we assume that any observable can be written as a sum of products of connected observables of less or equal order.

For a statistical mechanics problem with variables $\sigma_x$ on the sites of the lattice, an example of a two-point observable is the correlation $\langle \sigma_x \sigma_y \rangle$ induced by the Gibbs measure. The corresponding connected observable is $\langle \sigma_x \sigma_y \rangle-\langle \sigma_x  \rangle\langle  \sigma_y \rangle$.
For a system with quenched disorder an example of a two-point observable is $\overline{\langle \sigma_x \sigma_y \rangle}$. The corresponding connected observable is obtained 
subtracting the product $\overline{\langle \sigma_x\rangle} \overline{\langle\sigma_y \rangle}$ 
(with standard notation we indicate with $\langle\bullet\rangle$ thermal averages and with $\overline{\;\bullet\;}$ the average over the quenched disorder). 
Another two-point observable is  $\overline{\langle \sigma_x \rangle \langle \sigma_y \rangle}$.
For a percolation problem the simplest example of a  two-point observable  is the probability that the two lattice nodes $x$ and $y$ are occupied and are connected ({\it i.e.} they belong to the same cluster); similarly $n$-order functions in percolation are the probabilities that the $n$ nodes belong to the same cluster.

As we said before, in practice we assume that the problem we are dealing with is solvable on the Bethe lattice.
Similarly, we assume that the problem is also solvable (with a little bit more of technical effort) if the topology of the lattice is such that there are a few loops but otherwise the lattice extends to infinity in a tree-like fashion.

\begin{figure}
\begin{center}
\includegraphics[width=1\textwidth]{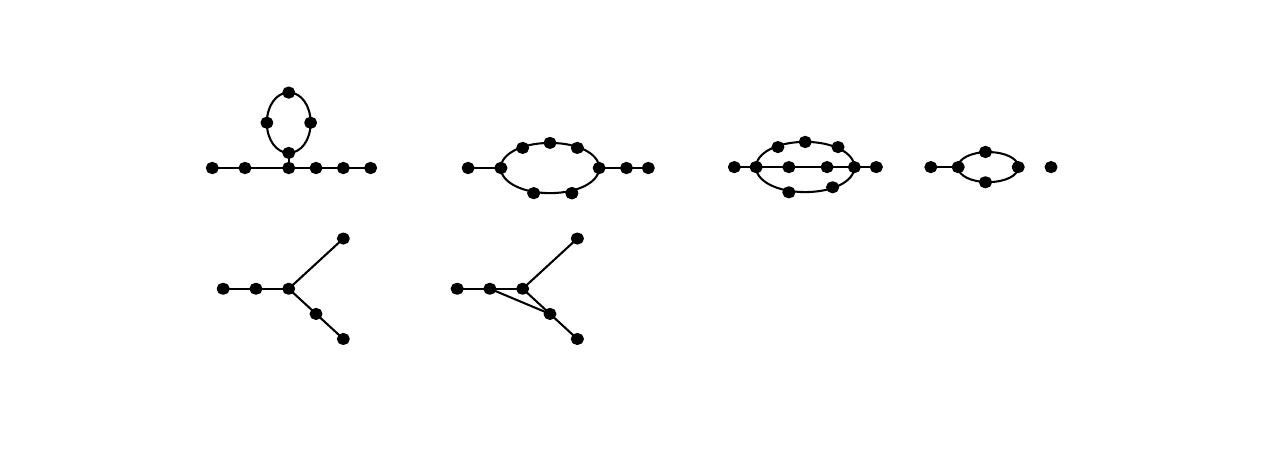}
\caption{ \label{fig:fat-diagrams} A few examples of fat-diagrams. First line: fat-diagrams with two external lines needed to evaluate two-point observables. 
Second line: fat-diagrams with three external lines needed to evaluate three point observables.
They give a vanishing or a finite contribution depending on the observable that we are looking at: in the first line
the first three give a finite contribution both to a two-point connected or disconnected correlation function while a connected correlation function vanishes on the last diagram.}
\end{center}
\end{figure}

If we have to evaluate an observable of order $n$ on such a lattice we can study independently the tree-like portions that extend to infinity. Graphically this amounts to prune all tree-like parts of the graph. After this process, we are left with a {\it finite} graph that contains the nodes $x_1,\dots,x_n $ and all the other nodes that were not removed in the pruning process because they are connected to at least {\it two} of the $x_1,\dots ,x_n $.
This structure is what we call a \textit{fat-diagram}. More precisely in our definitions a  fat-diagram of order $n$ is a graph with the following properties:
\begin{enumerate}
\item There are $n$ distinguished vertexes, which can have any degree (including zero or one). They are called \textit{external vertices}.
\item Other vertexes must have a degree equal or higher than two. We call them \textit{internal vertices}.
\item Each internal vertex must belong to the connected component of one of the internal vertices. 
\end{enumerate}
In other words, fat graphs correspond to elements of the set of graphs with $n$ external vertexes, no dangling edges and no fully disconnected components.
Furthermore, it is obvious that disjoint components of a fat-diagram can be studied independently and their properties can be expressed in terms of connected fat-diagrams. 
Examples of fat-diagrams are shown in Fig. \ref{fig:fat-diagrams}.

In general, however, the lattices we consider do not resemble at all a Bethe lattice.
Given a $n$-point observable $A$ depending on a $n$ vertex on the original lattice, we can identify a certain number of fat-diagrams $G$ with $n$ external legs that can be obtained by removing lines from the original graph.
{\it For a given fat-diagram $G$ we define the observable $A(G)$ as the value that the observable takes on a graph that has the structure of the fat-diagram and extends to infinity in a tree-like fashion that mimics the local structure of the original lattice}. 

\begin{figure}[h]
\includegraphics[width=1\textwidth]{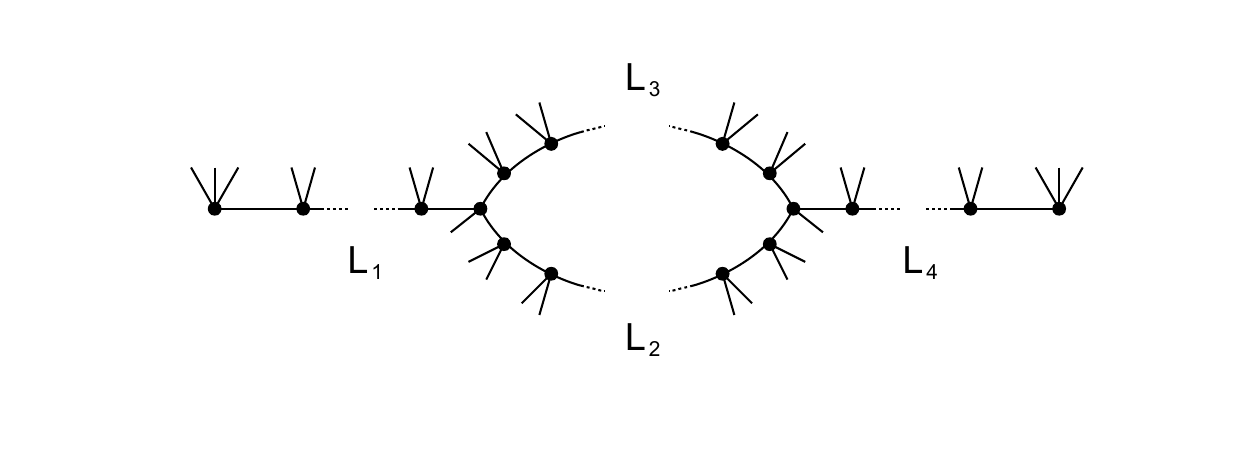}
\caption{When observables are computed over fat-diagrams one must attach to the sites as many infinite trees as needed to recover the connectivity of the original lattice. 
For instance when the second fat-diagram in Fig. \ref{fig:fat-diagrams} is evaluated on a  2D lattice, one should consider the graph displayed in the figure \ref{fig:fat-diagrams}
(where generic length $L_1$, $L_2$, $L_3$ and $L_4$ are assumed). 
Each spin of the fat-diagram has connectivity $2D=4$ and the additional lines represent the infinite trees whose contribution is the same of the Bethe lattice.}
\label{Fig:DiagramBethe}
\end{figure}

As an example, the structure associated to the second fat-diagram in Fig. \ref{fig:fat-diagrams} for a 
2D lattice is shown in Fig. \ref{Fig:DiagramBethe}. Each spin of the fat-diagram has connectivity $2D=4$ and its neighbors on the Bethe lattice are considered independent if the site is removed, exactly as in a Bethe lattice.

Given the observable $A(G)$, we now {\it define} a {\it line-connected} observable implicitly through the condition 
that the value of the observable on a fat-diagram $G$ is equal to the sum of the line-connected observables on all fat-diagrams $G'$ contained in $G$: 
\beq
A(G)= \sum_{G'\subseteq G}A_{lc}(G')
\label{exact}
\eeq
The inversion of Eq. \eqref{exact} can be written as
\beq
A_{lc}(G')=\sum_{G\subseteq G'} c_{G'\!,G} A(G),
\label{moebius}
\eeq
with appropriate coefficients $c_{G'\!,G''}$ that are discussed in the literature in the context of the so-called Moebius inversions in incidence algebra \cite{Stanley2011}. In practice line-connected observables can be determined iteratively starting from the simplest fat-diagrams (the trees) and considering fat-diagrams with increasing number of loops. To illustrate Eqs.(\ref{exact}) and (\ref{moebius}),  we display the result for the usual 2-points one-loop diagram: 
\begin{eqnarray}
\begin{split}
A \left( \feyn{fcf} \right) & = A_{lc} \left( \feyn{fcf} \right) + A_{lc}\left( \feyn{fc} \hspace{0.3cm} {\cdot} \right)  + A_{lc} \left({\cdot} \hspace{0.3cm} \feyn{cf} \right) + \\
& + A_{lc} \left(  \feyn{f} \hspace{0.5cm} \feyn{lS}  \hspace{0.5cm} \feyn{f}  \right) + A_{lc} \left(  \feyn{f} \hspace{0.45cm} \feyn{lSu} \hspace{0.45cm} \feyn{f}  \right) + A_{lc}\left( \cdot \hspace{0.3cm} \cdot \right)   \ .
 \end{split}
 \end{eqnarray}
Applying again Eq. (\ref{exact}) to the tadpole fat-diagram, we get:
\begin{eqnarray}
A \left( \feyn{fc} \hspace{0.3cm} \feyn{\cdot} \right) = A_{lc} \left( \feyn{fc} \hspace{0.3cm} \feyn{\cdot} \right) + A_{lc} \left( \cdot \hspace{0.3cm} \cdot \right)
 \end{eqnarray}
Collecting together these expressions above and the following identity condition:
\begin{eqnarray}
A \left (\cdot \hspace{0.3cm} \cdot \right) = A_{lc} \left( \cdot \hspace{0.3cm} \cdot   \right)
 \end{eqnarray}
we easily recover a compact expression for the line-connected observable, which turns out to be:
\begin{eqnarray}
\begin{split}
A_{lc} \left( \feyn{fcf} \right) = & A \left( \feyn{fcf} \right) -  A\left( \feyn{f} \hspace{0.5cm} \feyn{lS} \hspace{0.5cm} \feyn{f} \right)-A \left( \feyn{f} \hspace{0.45cm} \feyn{lSu} \hspace{0.45cm} \feyn{f} \right)\\
& -A \left( \feyn{fc} \hspace{0.3cm} \cdot \right)-A\left( \cdot \hspace{0.3cm} \feyn{cf} \right)+ 3 A \left( \cdot \hspace{0.3cm} \cdot \right)
\end{split}
\label{eq:M^2}
\end{eqnarray}
Note that if the observable $A$ is a connected one, the last three terms of the previous equation, associated to disconnected diagrams, vanish. For two-point observables, in the context of the M-Layer construction of Sec. \ref{sec:leadingorder} we have seen that the leading order is $O(1/M)$ and it is just the sum of the contribution of single lines, considered as independent.
The diagram in Eq. (\ref{eq:M^2}) contributes to the order $1/M^2$: we are considering a single loop. 

An important property of line-connected observables is that  they tend to zero if any line of the fat-diagram is very long, hence their name. This property will be proved later in section (\ref{Sec:lc}) using an explicit expression for them.


It is clear that the {\it exact} value of an observable on a given lattice is given $A(G)$ if $G$ coincides with the whole lattice. On the other hand, applying Eq.~(\ref{exact}) to the whole lattice amounts to write $A(G)$ as a  sum over all fat-diagrams of the original lattice. Note that the sum is finite if the lattice is finite.
If we order the sum over fat-diagrams according to the number of loops we thus obtain a graph theoretical loop expansion: this is the main result of this section.

Let us discuss the above results. The use of fat-diagrams seems rather arbitrary given that we do not know if the original lattice resembles or not a Bethe lattice.  
Actually, the exact result is obtained because, when we sum over all fat-diagrams $G'\subseteq G$, all the contributions by fat-diagrams $G'\neq G$ in the end cancels.
Nevertheless, we may ask what is the error if we sum only a subset of the fat-diagrams. 
The problem is that the contributions of different fat-diagrams appear to be of the same order of magnitude and it seems that neglecting some of them affects the results significantly.
More precisely we expect that the error is small only if we consider all possible fat-diagrams of size up to the correlation length. 
Thus all truncation schemes are bound to fail close to a critical point where the correlation length diverges.
Thus the graph theoretical expansion is exact but non-perturbative and in this sense is similar to the loop expansion in field theory.

In the next section instead we will show that the expansion is perturbative in the $M$-layer construction and $1/M$ is the small expansion parameter.
We stress the similarity with the fully-connected $M$-layer construction that makes the field-theoretical loop expansion perturbative.
Indeed in the case of the $M$-layer we can show that, although the exact result is only obtained summing all the series, if we consider a partial sum of all fat-diagrams with a number of loops up to $L$, the error decrease with $1/M^L$. 

At this point, the careful reader could object that in the case $M=1$ we are in trouble: we again need
to sum the whole series because the error is not decreasing with $L$. 
Nevertheless, we are not interested in the non-universal quantitative behavior of the various models but rather on their behavior close to a second order phase transition. In this perspective, we may invoke universality to claim that the universal quantities, {\it e.g.} critical exponents, do not depend on $M$.

\subsection{The Graph-Theoretical Expansion on the $M$-lattice}

In this subsection, we will show that the graph-theoretical loop expansion of the previous subsection is a perturbative expansion for the $M$-layer lattice, with the quantity $1/M$ playing the role of the small expansion parameter.

We have to evaluate a certain observable of order $n$ on the $M$-lattice averaged over all possible rewirings.
On each re-wiring of the $M$-lattice we can apply the exact formula (\ref{exact}), thus 
the average can be written as:
\beq
\langle A \rangle_{rew}= \sum_{G} P(G)\, A_{lc}(G)
\label{rewexact}
\eeq
The above expression means that we are summing over all possible fat-diagrams $G$ of the $M$-lattice, 
each weighted with the probability $P(G)$ that the given fat-diagram $G$ occurs over all possible rewirings. 

We note that the easiest way to understand the following arguments is to consider the case in which the original model is a two-dimensional lattice 
and the various $M$ layers can be though as an additional vertical coordinate as in Fig. (\ref{Mlayers}).

In order to proceed, it is convenient to classify all fat-diagrams on the $M$-lattice according to their projection on the original lattice.
The projections we consider retain all information on the topology and lengths of the original fat diagram, the only information that is lost is the specific layers over which it actually lives. As a consequence the projections are essentially fat diagrams that can be draw on the original $M=1$ lattice but can have multiple occupancies of sites and bonds.
The only constraints is that any two lines entering on one node from the same edge cannot be connected on either nodes.  For instance zero-loop fat diagrams are self-avoiding walks on the $M$-layer lattice, while their projections are non-backtracking walks on the original lattice $M=1$. 
Multiple occupancy is allowed for projections and removes the self-avoidance condition while the constraint imposes the non-backtracking property.
The use of projections is essential because  the value of the observable $A(G)$, and thus $A_{lc}(G)$, depends only on the topology and lengths of the fat diagrams and thus they are the same for all fat-diagrams with the same projection.

We recall that here we are not making any assumption of periodicity or homogeneity of the original lattice and therefore this result is totally general. We can thus write
\beq
\langle A \rangle_{rew}= \sideset{}{'}\sum_{G} W(G) A_{lc}(G)
\label{rewproj}
\eeq
where the prime means that the sum is in the space of projected fat-diagrams on the original lattice.
The weight $W(G)$ is the sum over all fat-diagrams of the $M$-lattice with vertical projection equal to $G$ each weighted with its probability $P(G)$.

This allows to split the problem into the computation of the observable $A(G)$ that does not depend on $M$, and of the term $W(G)$ that depends on $M$ and $G$ but not on the specific observable $A(G)$.
In the following we discuss the value of $W(G)$ showing that at leading order in $1/M$ it only depends on the number of loops in $G$, while the discussion of $A_{lc}(G)$ will be continued in section (\ref{Sec:lc}).

We will restrict the discussion of the weigt $W(G)$ to connected diagrams $G$ (see the first three examples in Fig. \ref{fig:fat-diagrams}), under the assumption that $A$ is a connected observable and therefore vanishes on a disconnected fat-diagram.
Let us first consider the case in which the projection is such that no site of the original lattice is occupied more than once. 
In this case, one can see that the sum over all possible realizations on the $M$ lattice times the corresponding probabilities is exactly $1/M^{L+n-1}$ 
where $L$ is the number of loops of $G$ and $n$ is the number of external vertexes of the graph. 
This weight corresponds to what we found in sec. \ref{sec:leadingorder} in the case $L=0$, $n=2$.

\begin{figure}[h]
\centering
 \includegraphics[scale=1]{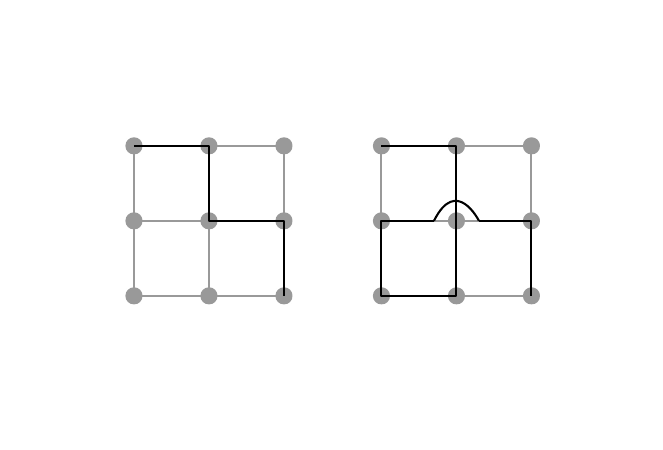} 
 \caption{
The $M$ dependence of the weight $W$ for a fat-diagram (black lines) is different if it do or do not have multiple occupancy points. 
For the left diagram we have $W=1/M$ exactly while for the right diagram we have $W=\frac{1}{M} \left( 1- \frac{1}{M} \right)$. 
The effect is subleading in powers of $1/M$ and is irrelevant for critical behavior.}
 \label{twopaths}
\end{figure}

Instead, if the projection is such that some sites on the original lattice are occupied more than once the corresponding factor is $1/M^{L+n-1}$ 
at leading order in $M$ but there are small corrections proportional to $1/M$. 
The presence of these corrections can be also understood noticing that for $M=1$ the probability 
of a projected fat-diagram with multiple occupancies must be zero.
 The above argument can be easily understood looking at Fig.~\ref{twopaths} where  the simple case of a fat-diagram with zero loops $L=0$ and $n=2$
external vertices (meaning the fat-diagram is just a line) is illustrated.
In the case of the simple line, left of Fig.~\ref{twopaths}, we have seen in Sec. \ref{sec:leadingorder} that according to a probabilistic approach we have a contribution $1/M$ to proceed from the beginning to the end of the line. 
In fact, calling $x_i$ the positions of the points on the line on the original lattice, $i=0,...,L$, 
starting from a chosen layer $\alpha$, one can choose one among the $M$ layers for all the internal points,
and once summed over all the possible realizations one obtains a factor 1,
and at the end with probability $\frac{1}{M}$ there is a link between $x_{L-1}$ and $x_L^{\beta}$, at fixed $\beta$.

On the right of Fig.~\ref{twopaths}) we see that the central point is occupied 2 times. Naturally the multiple occupations of the central point must correspond to different layers, otherwise the original fat-diagram on the $M$-lattice would have a loop. Therefore the graph on the right of Fig. (\ref{twopaths}) contributes  $\frac{1}{M} \left( 1- \frac{1}{M} \right)$.
The factor $\left( 1- \frac{1}{M} \right)$ is counting the probability of choosing two different layers for the two occupations of
the central point. The probability of this diagram is zero when $M=1$ as expected and it is smaller than the case where no multiple occupancies occur.

Following the same arguments above one can conclude that: for a general fat-diagram associated to a $n$-point observable ({\it i.e.} with $n$ external legs) and $L$ loops, at leading order we have the following simple expression:
\beq
W(G) \approx {1 \over M^{L+n-1}} \ .
\eeq
 while the exact expression of $W(G)$ at all orders in $M$ has complicated $1/M$ corrections that are irrelevant as long as we are interested in critical behavior.

At this point it is interesting to note that when one start to compute $1/M$ corrections  the need to consider line-connected observables is not evident. Given two nodes on the $M$-layer lattice one could consider the probability that they are connected by a line and this would give the $1/M$ leading term. Then one could consider the probability that they are connected by a structure with one loop $G$, say the usual 2-points one-loop diagram discussed in the previous section. In this case the observable will be given by the original $A(G)$ but one should take into account that when $G$ is present also the two linear fat diagrams are present. However they were already included in the leading order term (and computed wrongly under the assumption that the other line was not present) therefore their contribution should now be subtracted. Thus the correct $1/M^2$ result turns out to be equal to the line-connected expression $A_{lc}(G)$ and not to $A(G)$ alone.
These subtractions can be carried on easily at lowest order in $1/M$ but become extremely complicated at higher orders. The importance of the line-connected formalism is that instead it allows to carry on this procedure automatically and the natural objects to be considered for the $1/M$ expansion.

We note that the formalism can be also used to study finite size corrections on random regular graphs of size $N$.  In this case one should start from the exact expression (\ref{rewexact}) where $P(G)$ would be the probability over the random regular graph.
One should then sum $P(G)$ over all fat diagrams with the same topology and lengths. This amounts to consider all possible fat diagram in fig.~\ref{fig:fat-diagrams} summing over the possible 
labelling of their vertexes in terms of the $N$ nodes. At leading order in $1/N$ one can easily see that this contribute a factor 
\beq
{1 \over N^{L+n-1}} \ .
\eeq 
The analysis of critical behavior (non trivial finite-size effects in the critical region) can then be carried on following the same steps of the $M$-layer lattice that we are going to discuss in the following section.

\subsection{Critical Behavior} \label{SubSec:critical behavior M}

The results of the previous subsections are completely general and valid for any lattice. Nevertheless, we are mainly interested in the application of the formalism to critical phenomena and therefore we want to study the thermodynamic limit close to a phase transition.
In the thermodynamic limit by definition the original lattice is infinite and therefore we have an infinite sum over projected fat-diagrams.
Furthermore, we work under the assumption that the generic observable $A(G)$ (and thus $A_{lc}(G)$) is homogeneous. More precisely its value depends only on the topology of the graph $G$ and on the length of its internal lines but {\it not} on the actual way in which the (projected) fat-diagram is actually realized on the original lattice.
Note that this assumption does include disordered systems with quenched disorder provided the observable is defined as an average over the disorder.

It is thus natural to order the sum (\ref{rewproj}) according to the topology of the projected fat-diagram, {\it i.e.} in terms of Feynman diagrams. 
Besides, since at leading order all diagrams with the same topology have the same $W(G) \approx 1/M^{L+n-1}$, 
we can consider their sum at fixed internal distances ignoring possible multiple occupancies that would lead to sub-leading contributions.

Thus the contribution of a given Feynman diagram at fixed values of the lengths can be written as a sum 
over the position of the internal vertexes and the orientations of the lines entering into each vertex, 
of the product of the number of non-backtracking paths of the corresponding length of each line.

For instance, with reference to our reference exemplary diagram shown in Fig. \ref{Fig:DiagramBethe}, the sum of all projected fat-diagrams with internal lines $L_1,L_2,L_3$ and $L_4$ contributes the following term to the average of a two-point observable:
\beqa
&  & {1 \over M^2 S(G)} \sum_{x_2,x_3,\mu_1,\mu_4, \{ \mu_2,\nu_2, \eta_2\},\{ \mu_3,\nu_3, \eta_3\}}  b_{L_1}(x_2-x_1;\mu_1,\mu_2) b_{L_2}(x_3-x_2;\nu_2,\nu_3)
\nonumber
\\
& &\qquad\times\  b_{L_3}(x_3-x_2;\eta_2,\eta_3)
b_{L_4}(x_4-x_3;\mu_3,\mu_4) A_{lc}(L_1,L_2,L_3,L_4)
\eeqa
where $x_1$ and $x_4$ are the positions of external vertexes and $x_2$ and $x_3$ are the positions of internal vertexes of the fat-diagram.
Note the presence of the symmetry factor $S(G)$ in order to account for possible multiple counting in the summation over the position of the internal vertexes 
and over the orientations of the lines entering into each vertex.
The symmetry factor will be discussed  in Appendix
\ref{sec:symmetry_fact}, but the key point is that it is an intrinsic property of the Feynman diagram and 
therefore {\it it is the same used in the field theoretical loop expansions}.
The function $b_L(x,\mu,\nu)$
is the number of non-backtracking paths of length $L$ leaving the origin along direction $\mu$ and arriving at position $x$ along direction $\nu$. 
The sum over the directions entering on a given vertex, for example $\{ \mu_2,\nu_2, \eta_2\}$, implicitly requires the directions to be all different.

A detailed analysis of the function $b_L(x,\mu,\nu)$ is given in the Appendix \ref{app:nbw}. 
If we assume that the number of paths does not depend on the directions 
(which is actually an accurate approximation at large distances) 
we can write the contribution of the diagram as a sum over the position of the internal vertexes of the  
product of a term $b_L(x_i-x_j)$ for each line $(i,j)$. 
The sum over the choice of the internal directions at each vertex of degree $n$ 
provides a factor ${{2D}\choose {n}}n!$ counting the number of ordered $n$-uples out of the $2D$ possible directions.  
As above, with reference to the above fat-diagram, we have:
\vspace{0.2cm}

\begin{eqnarray}
\hspace{0.3cm} \feyn{fs} \hspace{0.7cm} \feyn{l lu}  \hspace{0.7cm} \feyn{fs} \hspace{0.8cm}
& = & {1 \over M^2 S(G)} \sum_{x_2,x_3}  (2D)^2 \left({ (2D)!\over (2D-3)!}\right)^2 b_{L_1}(x_1-x_2) b_{L_2}(x_2-x_3)
\nonumber \\
& & b_{L_3}(x_2-x_3)b_{L_4}(x_3-x_4) A_{lc}(L_1,L_2,L_3,L_4)
\label{solitoapprox}
\end{eqnarray}
\vspace{0.1cm}

To determine the total contribution of a Feynman diagram we have to sum over the lengths of its lines.
We will show that each Feynman diagram gives a contribution that diverges at the critical point and that the divergence is determined by the behavior of the corresponding fat-diagram when all lines are very large.
While the results obtained up to now depends only on the $M$-layer construction, to proceed further we need to specify the large distance behavior of $A_{lc}(G)$. This is a property of the actual problem we are studying,  {\it e.g.} a finite-temperature phase transition with or without quenched disorder, a zero-temperature phase transition, a percolation problem or any process that has a critical point on the Bethe lattice.

Later in sec. \ref{Sec:lc} we will discuss in more details the properties of $A_{lc}(G)$ , 
while in the following we will discuss why the large-distance behavior of $A_{lc}(G)$  determines the final result.
In the simplest case (realized {\it e.g.} in percolation)  $A_{lc}(G)$ can be approximated at leading order as the product over each line $L_{ij}$ of the fat-diagram, 
of a factor $\lambda^{L_{ij}}$ with $\lambda \rightarrow \lambda_c=1/(2D-1)$, as defined above, approaching the critical point. Again with reference to the diagram considered above we would have in this case:
\beq
A_{lc}(L_1,L_2,L_3,L_4) \approx c(G)  \lambda^{L_1+L_2+L_3+L_4}
\eeq
where the unspecified constant $c(G)$ depends on the model and determines  the critical exponents.
The above expression is only valid at large distances but it gives the correct critical behavior. Indeed if we insert it into (\ref{solitoapprox}) and sum over all internal distances $L_1,L_2,L_3$ and $L_4$ we obtain an expression that depends on the generating function of the non-backtracking paths $B_{\lambda}(x,y)\equiv \sum_{L=1}^\infty b_L(x,y)  \lambda^L $. 
Going to Fourier space  we arrive to an expression involving the Fourier transform  
$\hat{B}_{\lambda}(k)$ that has a singularity for $\lambda=\lambda_c$ and can be approximated at small momenta as
$\hat{B}_{\lambda}^{-1}(k) \propto \lambda-\lambda_c+k^2$ (see Appendix \ref{app:nbw}).
Again to be definite let us consider the usual diagram, whose contribution turns out to be:
\beq
{u^2 \over M^2 S(G)}{1 \over m^2+k^2}\left( \int d^D q {1 \over m^2+q^2}{1 \over m^2+(k-q)^2} \right){1 \over m^2+k^2}
\eeq
where $m^2 \equiv  \lambda-\lambda_c$ and $u$ is an effective (bare) coupling constant that absorbs all vertex contributions and possible rescalings.

We are now in the position to state the Feynman rules to determine the critical behavior of the $1/M$ expansion.
Given a connected observable of order $n$ we have to sum over all connected Feynman diagrams with $n$ external legs.
For each Feynman diagram one has to:
\begin{itemize}

\item multiply a factor $1/M^{n-1+L}$ where $L$ is the number of loops of the diagram.

\item divide by the symmetry factor $S(G)$ of the Feynman graph.

\item for each internal vertex of degree $m$ multiply by a factor $(2D)!/(2D-m)!$. 

\item Study  $A_{lc}(G)$ where $G$ is the fat-diagram with the topology of the Feynman diagram when the internal lines are large. 
As we said before, this is where the properties of the actual model enter into place. 
In the simplest case discussed above, we would have a factor $\lambda^L$ for each internal line. At this point we should multiply by the term $b_L$ and sum over the length of the lines $L$.
If the observable has the behaviour $\lambda^L$, 
it will result in multipling by a factor $1/(m^2+k^2)$ for each leg of the Feynman diagram.
In general we could have a more complicated situation. 
For instance, the contribution of a given line could be not just $\lambda^L$ but rather   $\frac{d^p}{d\lambda^p} \lambda^L$, 
in this case we should multiply a factor 
\beq
{d^p \over d\lambda^p}{1 \over \lambda-\lambda_c+k^2}
\eeq
for the corresponding line. This phenomenon with $p=1$ occurs in the random-field Ising model. 

\item perform the integration over the momenta that are not fixed by momentum conservation at the vertexes.
 
\end{itemize}

It should be clear that this analysis is only correct near a critical point and does not give the exact $1/M$ expansion but only the leading divergent part. Consistently, 
we expect that the degree of divergence increase with the number of loops and therefore the maximally divergent contribution at a given order in $1/M^p$ is given by the diagrams with $L=p+1-n$ loops. 
Consider the contribution to the two-point function induced by the non-backtracking paths. We have argued that this is $O(1/M)$ but has also subleading $1/M^2$ corrections.
We have neglected these corrections and assumed that at order $1/M^2$ the leading divergent term is given by the fat-diagrams with one loop.
Strictly speaking in order to justify this statement one should have a way to analyze sub-leading corrections induced by multiple occupancies of the nodes. Although we have not performed this analysis, we believe that the analysis is correct because the  zero-loop $1/M$ term gives a contribution that diverges less than the one-loop $1/M^2$ term, and therefore the terms coming from the $1/M^2$ corrections to the zero-loop terms should be irrelevant compared from those of the one-loop term.

\subsection{The expression for line-connected observables} \label{Sec:lc}

In section (\ref{SubSec:graph expansion}) we have given the implicit definition of line connected observables 
\beq
A(G)= \sum_{G'\subseteq G}A_{lc}(G')
\eeq
and we have shown how this expression can be used to compute algebraically $A_{lc}(G)$ for a given fat diagram starting from the simplest structures.
In the following we will provide instead an explicit expression that is essential for applications to critical phenomena and allows to prove the essential property of line-connected observables mentioned in section (\ref{SubSec:graph expansion}).
The explicit formula is the following:
\beq
A_{lc}(G)=\sum_{\widetilde{G}'\subseteq G}(-1)^{E(G)-E(\widetilde{G}')}A(\widetilde{G}')
\label{DEFOLC}
\eeq
where $E(G)$ is the number of edges in $G$. At variance with Eq. (\ref{moebius}), where the sum runs on (sub-)fat-diagrams, here the sum runs over  "tilded" sub-graphs $\widetilde{G}$ that {\it may also have dangling edges}, i.e. where some of the internal vertices can have degree one.
The validity of the above explicit expression can be proved through the following steps. First we note that the above expression  vanishes on a graph with at least one dangling edge: 
\beq
A_{lc}(\tilde{G}) \equiv \sum_{\widetilde{G}' \subseteq \widetilde{G}}(-1)^{E(\tilde{G})-
E(\tilde{G}')}A(\tilde{G}')=0 \ ,  \hspace{0.5cm} \ \mathrm{if} \ \widetilde{G} \neq G \ .
\eeq
This is due to the fact that $A(\tilde{G})$ is equal to $A(G)$ where $G$ is the graph obtained from $\tilde{G}$ removing all dangling edges.
Therefore we can write:
\beq
A(G) = \sum_{G'\subseteq G}A_{lc}(G')=\sum_{\widetilde{G}'\in G}A_{lc}(\tilde G')
\eeq
and then we can use the following identity to prove the formula (\ref{DEFOLC}) for the line-connected observable:
\beq
\sum_{\tilde{G}' : \, \widetilde{G}'' \subseteq \widetilde{G}' \subseteq \widetilde{G} }(-1)^{E(\widetilde{G}')-E(\widetilde{G}'')}=\delta(\widetilde{G},\widetilde{G}'')
\eeq
where $\delta(\widetilde{G},\widetilde{G}'')$ is  equal to one if $\widetilde{G}=\widetilde{G}''$ and zero otherwise.
Note that the sum over all tilded $\widetilde{G}'$ that are equal to a fat-diagram $G'$ once their dangling edges are removed gives the factor $c_{G,G'}$ in the Moebius inversion formula.

The interpretation of the explicit formula (\ref{DEFOLC}) is straightforward: in order to compute $A_{lc}(G)$, we have to evaluate the observable $A$ on each of the subgraphs that are obtained from $G$ by removing sequentially its edges times a factor $-1$ for each edge removed.
This result (and the explicit formula (\ref{DEFOLC})) can be further simplified considering all edges on a given line. One can easily see that the effect of applying the procedure to all edges of a given line is equivalent to evaluate the observable in presence of the line and subtract the observable when the line is removed.  
This leads to the result that {\it in order to compute $A_{lc}(G)$, we have to evaluate the observable $A$ on each of the subgraphs that are obtained from $G$ by removing sequentially its lines times a factor $-1$ for each line removed}.

We can now discuss the main difference between $A(G)$ and $A_{lc}(G)$. It lies in their behavior when one of the lines in $G$, 
say $l$, tends to infinity. In this case, $A(G)$ tends to a constant: the value that it has on the subgraph where $l$ 
has been removed. On the other hand, $A_{lc}(G)$ can be written as the sum of certain diagrams {\it with} $l$ 
minus the same expression evaluated on diagrams {\it without} $l$, and since the first expression tends to the second one when the $l$ tends to infinity, it follows that $A_{lc}(G)$ tends to zero.
This is also why we use the name {\it line-connected} observables.

To understand the importance of line-connectedness we recall that in the previous section we have ordered the sum over fat-diagrams collecting all fat-diagrams with the same topology (the Feynman diagrams) and summed over all possible internal lengths. 
If $A_{lc}(G)$ would tend to a constant when one of its internal lines tends to infinity,
the procedure will be in trouble because the contribution of the Feynman diagram would be also infinite.

In the previous subsection, we have shown that critical behavior is determined by the behavior of the line-connected observables at large distances. This is a specific property of the model and no general results can be obtained. 
In the following, we will discuss the case of a statistical mechanics problem at finite temperature. Here considerable progress can be made and one can show that, 
as expected on grounds of universality, the universal predictions obtained through the graph theoretical loop expansion are the same as those of the field theoretical loop expansion
(provided that the physics on the Bethe lattice is the same as on the FC model).

In order to be definite let us consider a statistical mechanics problem where on the nodes of the lattice there are real variables 
$\sigma_i$, as in Section \ref{sec:leadingorder}. Since the notation is cumbersome, we shall drop the apex in $P^{\tB}$ used over there. At this point the reader should be able to understand when quantities are computed on the infinite Bethe lattice.
The most general two-point observable is the probability $P(\sigma,\tau)$ defined according to the Gibbs distribution of the problem.
In this case the computation of the $A(G)$ amounts to: 
i) multiply a term $Z_L(\sigma_i,\sigma_j)$ for each line of length $L$ connecting vertexes $i$ and $j$; 
ii) multiply a term $\rho_{2D-n}(\sigma)$ for each vertex of degree $n$; 
iii) sum over the configuration of the internal vertexes; iv) normalize the result. Here and in what follows  
$Z_L(\sigma_i,\sigma_j)$ is a non-normalized probability distribution on a one-dimensional line where the two end-points have degree one. As we discuss in Appendix\ref{app:bethe}, we can express $Z_L$ in the form
\beq
Z_L(\sigma,\tau) = Q_{L}(\sigma)Q_{L}(\sigma) + \sum_{|\lambda|<1} \lambda^L g_\lambda(\sigma)g_\lambda(\tau),
\eeq
where $Q(\sigma)$ is the Bethe cavity distribution and  the functions $g_\lambda(\sigma)$ can be associated to the eigenvectors of a transfer matrix.

The functions $\rho_k(\sigma)$ instead are the (non-normalized) marginal distributions of a variable in the Bethe lattice when $2D-k$ of its incident edges are removed from the graph (see again Appendix \ref{app:bethe}):
\begin{equation}
\rho_k(\sigma)= Q^k(\sigma)\, e^{\beta H(\sigma)},
\label{rhok}
\end{equation}
where $H(\sigma)$ represents an external field. For convenience, in what follow we will assume $\Sigma_\sigma$ to denote an integration $\int d\mu(\sigma)$, where $d\mu(\sigma)$ is the single variable prior distribution of our problem.

With reference to the diagram discussed above in Fig. \ref{Fig:DiagramBethe}, 
we have the following expression for the observable $A=P(\sigma_1,\sigma_4)$:
\beq
A(G)={1 \over \mathcal{N}(G)}\sum_{\sigma_2,\sigma_3}\rho_{2D-1}(\sigma_1)Z_{L_1}(\sigma_1,\sigma_2)\rho_{2D-3}(\sigma_2)
Z_{L_2}(\sigma_2,\sigma_3)Z_{L_3}(\sigma_2,\sigma_3)\rho_{2D-3}(\sigma_3)Z_{L_4}(\sigma_3,\sigma_4)\rho_{2D-1}(\sigma_4)
\eeq
where the normalization $\mathcal{N}(G)$ reads: 
\beq
\mathcal{N}(G)=\sum_{\sigma_1,\sigma_2,\sigma_3,\sigma_4}\rho_{2D-1}(\sigma_1)Z_{L_1}(\sigma_1,\sigma_2)\rho_{2D-3}(\sigma_2)
Z_{L_2}(\sigma_2,\sigma_3)Z_{L_3}(\sigma_2,\sigma_3)\rho_{2D-3}(\sigma_3)Z_{L_4}(\sigma_3,\sigma_4)\rho_{2D-1}(\sigma_4) \ .
\eeq
In order to compute $A_{lc}(G)$ we have to remove sequentially all the lines in $G$. Computing $A$ on the diagram without a line $L_k$ amounts to evaluate the above expressions with the replacement
\beq
Z_{L_k}(\sigma,\tau) \rightarrow Q(\sigma)Q(\tau) \ .
\eeq
Here $Q(\sigma)$ is the marginal distribution for a variable in the Bethe lattice when all but one of its incident edges have been pruned from the graph (as we discuss Appendix \ref{app:bethe}).
As we saw in the previous section, we are interested in the behavior of $A_{lc}(G)$ when all lines are large.
This is controlled by the behavior of $Z_{L}(\sigma,\tau)$ for large $L$. Let us assume that we are in the simplest case in which we have in the large $L$ limit
\beq
Z_{L}(\sigma,\tau) \approx Q(\sigma)Q(\tau)+\lambda^{L}g(\sigma)g(\tau),
\label{PLapprox}
\eeq
where $\lambda$ is the critical eigenvalue, equal to $\lambda_c=1/(2D-1)$ precisely at the critical point, and 
$g(\sigma)$ is the associated eigenfunction.
We can now see that the denominator of any tilded subdiagrams $\widetilde{G} \in G$ is given at leading order by:
\beq
\mathcal{N}(\widetilde{G}) \approx \left(\sum_\sigma \rho_{2D}(\sigma)\right)^4 \ ,
\eeq
where the exponent four comes from the number of vertexes.
Since the denominator of $A(\widetilde{G})$ is the same at leading order, we can now collect all numerators of the diagrams involved in $A_{lc}(G)$. 
Now the peculiar form of the expression (\ref{DEFOLC}) comes to place: since it requires that we sum over all lines with a factor $-1$ for a removed line, we can write
\beqa
A_{lc}(G)&\approx &{1 \over \left(\sum_\sigma \rho_{2D}(\sigma)\right)^4}\sum_{\sigma_2,\sigma_3}\rho_{2D-1}(\sigma_1)\tilde Z_{L_1}(\sigma_1,\sigma_2)\rho_{2D-3}(\sigma_2)
\tilde Z_{L_2}(\sigma_2,\sigma_3)
\nonumber
\\
& & \qquad\times \tilde Z_{L_3}(\sigma_2,\sigma_3)\rho_{2D-3}(\sigma_3)\tilde Z_{L_4}(\sigma_3,\sigma_4)\rho_{2D-1}(\sigma_4)
\eeqa
where 
\beq
\tilde Z_{L}(\sigma,\tau) \equiv Z_{L}(\sigma,\tau) -  Q(\sigma)Q(\tau) \approx \lambda^{L}g(\sigma)g(\tau)
\eeq
leading to:
\beq
A_{lc}(G)\approx  {  \rho_{2D-1}(\sigma_1)g(\sigma_1)\rho_{2D-1}(\sigma_4)g(\sigma_4)  \left(\sum_\sigma \rho_{2D-3}(\sigma)g^3(\sigma)\right)^2\over \left(\sum_\sigma \rho_{2D}(\sigma)\right)^4} \lambda^{L_1+L_2+L_3+L_4} \ .
\eeq
This result can be generalized to higher order connected correlations and more Feynman diagrams involving vertexes of all degrees. 
It implies that for such a problem, in order to compute the contribution of a given Feynman diagram, once we have summed over 
the length of all lines, we would have to multiply by a factor 
\begin{equation}
\frac{\sum_\sigma \rho_{2D-n}(\sigma)g^n(\sigma)} {\sum_\sigma \rho_{2D}(\sigma)}
\ \ \ \ \mbox{and} \ \ \ \  
\frac{\rho_{2D-n}(\sigma)g^n(\sigma)} {\sum_\sigma \rho_{2D}(\sigma)}\equiv a^{(n)}(\sigma)
\end{equation}
respectively for each internal and external vertex of degree $n$. 

One finds out that the resulting contribution is divergent
exactly at the critical point where $\lambda\rightarrow\lambda_c$. 
This holds for any Feynman diagrams of any connected correlations, provided that the contraction of the observable with $a^{(n)}(\sigma)$ is not zero.
We conclude that the critical behavior of such a theory is completely equivalent to a scalar cubic theory, as required by universality since the symmetry factors are the same.

The above results can be easily generalized to more complex situations. 
In general, for any phase transition in the universality class of a given field theory, one should be able 
(maybe with some difficulty) to recover the very same field theory from the $1/M$ expansion. 
Much more interesting are those cases in which no field-theoretical mapping is known, 
these include notably disordered systems at zero temperature and various percolation problems.

In these more complex cases, one should evaluate $A_{lc}(G)$ explicitly. In case $A$ is a connected observable, one should only consider connected subgraphs. For instance, with reference to the diagram considered above, one should sum $A$ evaluated on the original diagram and subtract its value on the diagrams with respective lines $L_2$ and $L_3$ missing.

It is also instructive to check how the above mapping to a scalar $\phi^3$ theory behaves for a ferromagnetic transition, 
in which we expect instead a mapping to a $\phi^4$ theory. 
The key point is the internal vertex factor
\begin{equation}
V\equiv\frac{\sum_\sigma \rho_{2D-3}(\sigma)g^3(\sigma)} {\sum_\sigma \rho_{2D}(\sigma)}\,.
\end{equation} 
It turns out that in the case of a ferromagnetic transition $Q(\sigma)$ is even as a function of $\sigma$ and so are the functions $\rho_k(\sigma)$ defined by  Eq. \eqref{rhok},  while the critical eigenvector $g(\sigma)$ is odd.
Therefore the cubic vertex factor $V$ is zero. 
In order to get a finite contribution we cannot attach three terms $\lambda^L g(\sigma)g(\tau)$ to a cubic vertex and we have to consider the less divergent terms in the expansion (\ref{PLapprox}). 

This implies that at least for one of the legs of the vertex we should take the term associated to the second eigenvalue
$\lambda_2^L g_2(\sigma)g_2(\tau)$ that gives a non-divergent contribution because $\lambda_2<\lambda_c$ (supposing that we are considering a model without degeneration of the
eigenvalues at the critical point).
For this leg, the largest terms are those associated with small length $L$,
and this acts as a contraction of two vertexes of degree three into a single vertex of degree four. In the second diagram in Fig. \ref{fig:fat-diagrams} 
we have to assume that one leg (either $L_2$ or $L_3$) remains finite and in the first diagram in Fig. \ref{fig:fat-diagrams} 
the single line connecting the two internal vertexes must also remain finite, 
thus both diagrams are equivalent to the diagram with one loop and one quartic vertex that gives the first correction to the self-energy in the $\phi^4$ theory.  
As in the case of the $1/M$ expansion around the fully connected case of sec. \ref{FCMlayer}, also in the case
of the Bethe $M$-layer we recover the usual $\phi_4$ theory: universality is recovered as it should.

As a last illustration of the approach, we discuss site percolation. 
We consider the two-point observable that gives the probability that two points on the lattice are connected.
Calling $p$ the probability to have a link between two point nearest neighbors, with reference to  diagram of Fig.~\ref{Fig:DiagramBethe} we have that 
\beq
A \left( \feyn{fcf} \right)= p^{L_1+1} \, (1-(1-p^{L_2-1})(1-p^{L_3-1})) \, p^{L_4+1}= p^{L_1+1} \, (p^{L_2}+p^{L_3}- 2 p^{L_2+L_3-1}) \, p^{L_4}
\eeq
\beq
A \left( \feyn{f} \hspace{0.5cm} \feyn{flS} \hspace{0.5cm} \feyn{f} \right)=p \, p^{L_1} \, p^{L_2} \, p^{L_4}
\eeq
\beq
A \left( \feyn{f} \hspace{0.45cm} \feyn{lSu} \hspace{0.45cm} \feyn{f} \right) =p \, p^{L_1} \, p^{L_3} \, p^{L_4}
\eeq
and therefore the line-connected contribution is given by:
\beq
A_{lc} \left( \feyn{fcf} \right) =-2 \, p^{L_1+L_2+L_3+L_4}
\eeq

The above results show that, at variance with the ferromagnetic transition, the diagram is relevant, in agreement with the fact that percolation is associated with a cubic field theory. Extending the computation of the two-point function to highest orders in the loop expansion and supplementing it with the computation of the three-point function one can extract the critical exponents. This result is interesting because the series would be obtained without resorting to the   Fortuin-Kasteleyn mapping to the $q$-state Potts model, that requires a continuation to the $q=1$ case.

\subsection{The role of spurious disorder}
\label{sec:spurious}

In the previous section, we have shown that at criticality the Bethe $M$-layer loop expansion for the Ising model is the same of the $\phi^4$ theory.
This result is satisfying but requires some additional comments.
Indeed one should note that, at variance with the FC $M$-layer construction, the Bethe $M$-layer introduces always some spurious quenched disorder in the problem due to the fluctuations over the rewirings.
From the point of view of critical phenomena this kind of disorder may or not may be a relevant perturbation \cite{Cardy1996}. In the latter case, we should observe a different critical behavior when $M$ is different from one with respect to the $M=1$ case.
This is something we would like to avoid because we do not want the $M$-layer construction to change the universality class of the problem. On the other hand, above we selected the leading divergences at all orders in $1/M$ and the theory turned out to be equivalent to the pure theory, so where did disorder go?

A possible answer is the following. The disorder is essentially absent in the large $M$ limit because the Bethe lattice looks regular and homogeneous at any finite distance. A side effect is that annealed and quenched averages over the rewiring are strictly equivalent at $M=\infty$. 
We have argued before that at any large but {\it finite} $M$ the critical behavior of the pure model is observed in the (tiny) critical region (scaling as $M^{-2/\epsilon}$ 
according to sec. \ref{FCMlayer}).  Now disorder appears very weakly decreasing $M$ in such a way, in order to see the expected deviations from the critical behavior of the pure model, 
we have to zoom in a region smaller than  $M^{-2/\epsilon}$.
Therefore if we send $M$ to infinity while remaining at distance $O(M^{-2/\epsilon})$ from the critical point, the region where deviations from the pure critical behavior are observed shrink to zero on the scale $O(M^{-2/\epsilon})$ and this explains why disorder does not show up in the loop expansion.
This argument implies that  $M$-layer construction can be used to study the universality class of the pure model provided we take the $M \rightarrow \infty$ limit while studying a region of size $O(M^{-2/\epsilon})$ around the Bethe critical point.
On the other hand, this should suggest caution in studying numerically system at finite $M$ because the critical behavior could be different from the $M=1$ case if the disorder is a relevant perturbation. 

This scenario is suggested by a phenomenon that was studied some years ago in MF SG.
Consider the ensemble of random-regular-graphs (RRG) of size $M$. It is knows that fluctuations $\sigma_M$ of the ground state energy of SG models defined on the ensemble of random-regular-graphs (RRG) of size $M$ is $O(M^{-1/2})$ if the couplings $J_{ij}$ have a Gaussian distribution while they are  much smaller if the couplings have a bimodal distribution \cite{Parisi2010a,Bouchaud2003a}.
This is because, when the couplings are bimodal, the problem can be transformed locally by means of a gauge transformation in a ferromagnetic problem with constant $J$
and the lattice appears to be the same for each sample and therefore sample-to-sample fluctuations are much smaller.
Similarly, in the $M$-layer construction, it is true that some disorder must be induced by the random permutation, 
but locally the lattice is regular and we do not see these fluctuations as soon as $M$ is large enough.

\section{Conclusions}\label{Conclus}

Given a physical system defined on a generic lattice or factor graph, we introduced the Bethe $M$-layer construction as a way to build a different model that is exactly solved by the Bethe approximation in the large $M$ limit.
We have discussed the problem of computing $1/M$ corrections to the leading order Bethe result and showed that they
can be expressed in terms of a perturbative diagrammatic loop expansion in terms of so-called fat-diagrams.

Our motivation is to study some important second-order phase transitions that do exist on the Bethe lattice but are either different or absent in the corresponding fully connected case.
On physical grounds, we expect that when the construction is applied to a lattice in finite dimension there is a small region of the external parameters close to the Bethe critical point where deviations from mean-field behavior will be observed. In this region, the $1/M$ expansion for the corrections diverges at all orders and
it can be re-summed as usual to determine the correct non-mean-field  critical exponents.  

While the exact computation of the $1/M$ correction is rather complicated we have shown that in the critical region
 considerable simplification occurs. In the end, the critical series for the generic multi-point observable can be expressed as a sum of Feynman diagrams with the same numerical prefactors they have in field theories. 
The only difference with the field theoretical loop expansion is that the contribution of a given diagram must not be evaluated as usual associating Gaussian propagators to its lines and integrating over the vertex position.
Instead one should consider the graph as a portion of the original lattice attaching one-dimensional chains to the lines and the appropriate number of Bethe trees to each point in order to restore the connectivity of the original lattice.

The actual contribution of each (fat)-diagram is the so-called line-connected observable that includes also contributions from sub-diagrams with appropriate prefactors. 
Thus in practice, Feynman diagrams with their symmetry factors can be borrowed directly from field theories and one only needs to compute the line-connected observable 
of the corresponding fat-diagram in the limit of all lines becoming large.
Applications of the formalism to the RFIM at zero temperature and to the SG in a field at zero temperature are currently under investigation.

\subsection*{Acknowledgements} 
This work was developed over the course of the last seven years and we benefited from conversions with many colleagues, we wish to acknowledge in particular discussions with U. Ferrari and M. Ohzeki.
We acknowledge funding from the European Research Council
(ERC) under the European Union’s Horizon 2020 research and innovation
programme (grant agreement No [694925]).

\appendix

\section{Hamiltonian Systems on the Bethe Lattice}
\label{app:bethe}

In this section we recall some standard techniques that can be used on the Bethe lattice, the infinite tree where each vertex has connectivity $c$. What follows applies also in the thermodynamic limit  to models defined on random regular graphs, and, with some little generalization, to inhomogeneus locally tree-like graphs (e.g. Erd\'os-Renyi random graphs).
The Hamiltonian of a general model with one and two-body interactions  can be written as:
\beq
\mathcal{H}(\{\sigma_i\}) = -\sum_{(i,j)} J(\sigma_i,\sigma_j)- \sum_i H(\sigma_i)
\eeq
The present treatment is valid for any type of variables $\{\sigma_i\}$, not just Ising spins. We assume a factorized prior on the variables, $\prod_i d\mu(\sigma_i)$. The Ising case corresponds to $d\mu(\sigma_i) =\delta(\sigma_i+1)+\delta(\sigma_i-1)$.

The Hamiltonian is homogenous, in the sense that  all the one and two body terms in the Hamiltonian are the same. System with quenched disorder can be also discussed in this framework by means of the replica method. In this case the variable $\sigma_i$ are actually replicated variables, but once disorder average is taken the Hamiltonian is homogenous.

Under some ergordicity assumption on the Gibbs measure (namely Replica Symmetry \cite{Parisi1987}), 
the thermodynamic of the model can be solved once the solution $Q(\sigma)$, which we assume to be unique, of the following implicit equation is found (usually by recursion):
\beq
Q(\sigma) = \frac{1}{\mathcal Z_Q}\int d\mu(\tau)\ e^{\beta J(\sigma,\tau) +\beta H(\tau)}\ Q(\tau)^{c-1}
\label{recQ}
\eeq
where $c$ is the degree of connectivity and
\begin{equation}
\mathcal{Z}_Q = \int d\mu(\sigma)\, d\mu(\tau)\ e^{\beta J(\sigma,\tau) +\beta H(\tau)}\ Q(\tau)^{c-1}
\end{equation}
We call $Q(\sigma)$ the cavity distribution and we have $\int d\mu(\sigma)\, Q(\sigma)=1$ by construction.

The marginal probability distribution of a variable in the infinite Bethe lattice is then given by 
\beq
P^{\tB}(\sigma) \propto Q(\sigma)^c \, e^{\beta H(\sigma)}
\eeq
In order to compute two-point correlations among variables at finite distance $L$ on the Bethe lattice, we notice that they are exactly the same of the correlations on a one-dimensional system (of length $L$) with the same couplings $J(\sigma,\tau)$ and an effective  field on the internal variables due to $c-2$ cavity fields in addition to $H$. 
Therefore, to solve the one dimensional model,  we introduce the (unnormalized) cavity fields
\beq
\rho_{k}(\sigma) \equiv Q(\sigma)^{k}\,e^{\beta H(\sigma)}.
\eeq
and the (symmetric) transfer matrix
\beq
T(\sigma,\tau) \equiv \sqrt{\rho_{c-2}(\sigma)}\, e^{ \beta J(\sigma,\tau)}\, \sqrt{\rho_{c-2}(\tau)} 
\eeq

The transfer matrix can be written in terms of its orthonormal eigenvectors as
\beq
T(\sigma,\tau)= \sum_\lambda  \, \lambda\, e_\lambda(\sigma) e_\lambda (\tau),
\eeq
with $\int d\mu(\sigma)\, e_\lambda (\sigma) e_\gamma (\sigma) = \delta_{\lambda,\gamma}$.
Using Eq. \ref{recQ} one can see that the quantity
\beq
\sqrt{\rho_{c-2}(\sigma)}\,Q(\sigma)
\eeq
is actually an eigenvector of the transfer matrix. From the Perron-Frobenius theorem we can argue that it is the eigenvector corresponding to the largest eigenvalue $\lambda_{max}=\mathcal Z_Q$. Therefore we have
\beq
e_{\lambda_{max}} \propto \sqrt{\rho_c(\sigma)}
\eeq
For later convenience we call the above eigenvector the Bethe eigenvector:
\beq
e_{B}(\sigma) \equiv e_{\lambda_{max}}(\sigma)
\eeq
from which we immediately have that the marginal probability is given exactly by the square of 
\beq
P^{\tB}(\sigma)= e_{B}^2(\sigma)
\eeq
where the normalization of the probability follows from the normalization of the eigenvector.

In order to compute the two-point correlation we consider a closed one-dimensional chain of length $N$ and we then take the thermodynamic limit $N \rightarrow \infty$.
One easily obtains
\beq
P_L^{\tB}(\sigma,\tau)=\frac{ \sum_{\lambda,\gamma}\gamma^{N-L}\lambda^L e_\gamma(\sigma)e_\lambda(\sigma)  e_\gamma(\tau)e_\lambda(\tau)  }{\sum_{\lambda} \lambda^N}
\eeq
in the thermodynamics limit both terms proportional to $N $ are dominated by the largest eigenvalue and we have 
\beq
P_L^{\tB}(\sigma,\tau)=\sum_{\lambda}\left({\lambda \over\lambda_{max}} \right)^L e_B(\sigma)e_\lambda(\sigma)  e_B(\tau)e_\lambda(\tau)  
\eeq
Introducing the rescaled eigenvalues 
\beq
\frac{\lambda}{\lambda_{\max}} \rightarrow \lambda
\eeq
and defining the functions
\beq
a_{\lambda}(\sigma) \equiv e_B(\sigma)e_\lambda(\sigma) 
\eeq
we arrive to the form 
\beq
P_L^{\tB}(\sigma,\tau)=P^{\tB}(\sigma)P^{\tB}(\tau)+\sum_{|\lambda|<1} \lambda^L a_\lambda(\sigma) a_\lambda(\tau)  
\label{defPL}
\eeq
Note that we have $\int d\mu(\sigma)\, a_\lambda(\sigma)=0$ for all $\lambda$ smaller than one.

From the exact expression of the two-point distribution one can obtain two point susceptivities on the Bethe lattice
\beq
\chi(A) \equiv\sum_{i\neq 1}\, \bigg(\langle\, A(\sigma_1)A(\sigma_i)\, \rangle - \langle A(\sigma_1)\rangle\,\langle A(\sigma_i) \rangle \bigg)
\eeq 
as
\begin{equation}
\begin{aligned}
\chi(A) &= c\,\sum_{L=1}^{\infty} (c-1)^{L-1} \sum_{|\lambda|<1} \lambda^L \left( \int d \mu(\sigma)\ a_\lambda(\sigma) A(\sigma)\right)^2 \\
&=c\sum_{|\lambda|<1} {\lambda \over 1- \lambda/\lambda_c}\left( \int d \mu(\sigma)\ a_\lambda(\sigma) A(\sigma)\right)^2
\end{aligned}
\end{equation}
where
\beq
\lambda_c \equiv {1 \over c-1}\, .
\eeq
If the  largest  eigenvalue smaller than one approaches the critical value $\lambda_c$ the susceptivity diverges, provided that $A(\sigma)$ has a non-zero projection on the corresponding $a_{\lambda}(\sigma)$.

Let us now consider the one dimensional chain used in Sec. \ref{Sec:lc}. In such chains we have the contributions of $c-2$ cavity fields on the internal variables but insert no extra field at the extremities. We can write an effective partition function for this system, conditioned on the values of the spins at the extremities, $Z_L(\sigma,\tau)$, in terms of our transfer matrix $T$. We obtain 
\beq
Z_L(\sigma,\tau) = Q_{L}(\sigma)Q_{L}(\sigma) + \sum_{|\lambda|<1} \lambda^L g_\lambda(\sigma)g_\lambda(\tau)
\label{defZL}
\eeq
where 
\beq
g_\lambda(\sigma) \equiv \frac{e_\lambda(\sigma)\rho^{-1/2}_{c-2}(\sigma) }{\int d\mu(\sigma')\ e_B(\sigma')\rho^{-1/2}_{c-2}(\sigma')}.
\eeq

It is interesting to note that only the exact two-point function $P^\tB(\sigma,\tau)$ is somehow special, in the sense that other two-point distribution, such as the one we could obtain normalizing $Z_L(\sigma,\tau)$ to one, would contain a denominator of the form $\sum_{\lambda} \lambda^L  \left(  \int d\mu(\sigma) g_\lambda(\sigma)   \right)^2$ that depends explicitly on all eigenvalues. In fact $\int d\mu(\sigma)\, g_\lambda(\sigma) $ does not generally  vanish, while $\int d\mu(\sigma)\, a_\lambda(\sigma)=0$.

All the results presented in this Appendix can be rederived using a slightly approach, through a rescaled and non-symmetric transfer matrix:
\beq
\tilde T(\sigma,\tau) \equiv \frac{1}{\mathcal Z_Q}\,  e^{\beta J(\sigma,\tau)+\beta H(\tau)}\, Q(\tau)^{c-2}. 
\eeq
The leading right eigenvector of $\tilde T$ is $Q(\sigma)$, with associated eigenvalue $\lambda=1$, and the other eigenvalues have modulus smaller then one. So $\tilde T$ has the same eigenvalues of $T$ after the rescale.
Moreover the functions $g_\lambda(\sigma)$ happen to be the right eigenvector of $\tilde T$. The corresponding left eigenvectors shall be given by $\rho_{c-2}(\sigma)g_\lambda(\sigma)$, and the orthonormalization condition reads
\begin{equation}
 \int d\mu(\sigma)\,\rho_{c-2}(\sigma)g_\lambda(\sigma)g_\gamma(\sigma)=\delta_{\lambda,\gamma}.
\label{normg}
\end{equation}
 
In terms of $\tilde T$, we easily recover the expressions \eqref{defZL} and \eqref{defPL} for the open chains through the relations
\begin{equation}
Z_L(\sigma,\tau)= \tilde T^L(\sigma,\tau)\rho^{-1}_{c-2}(\tau); \qquad P^\tB_L(\sigma,\tau) = 
\rho_{c-1}(\sigma)\,
Z_L(\sigma,\tau)\, \rho_{c-1}(\tau)
\end{equation}
with $a_\lambda(\sigma)$ and $g_\lambda(\sigma)$ linked by
\begin{equation}
a_\lambda(\sigma)=\frac{g_\lambda(\sigma)\rho_{c-1}(\sigma)}{\int d\mu(\sigma')\,\rho_c(\sigma')}.
\end{equation}

\section{Non-Backtracking Walks}
\label{app:nbw}
We shall extend here some results of Parisi and Slanina and of Fitzner and van der Hofstad \cite{Fitzner2013} on the number of non-backtracking walks (NBW) on the lattice $\bbZ^{d}$, for $d \geq 2$ (the case $d=1$ is trivial).

We shall first introduce some conventions that we shall adopt throughout this Section. We shall index with $\mu,\nu$ and $\alpha$ the $2d$ directions in the $d$-dimensional lattice, each index taking values in $\{\pm 1,\pm 2, \dots,\pm d\} $. The corresponding versors $e_{\mu}\in \bbZ^d$ are defined by $(e_{\mu})_m\equiv \sign(\mu)\, \delta_{|\mu|,m}$ for $m=1,\dots,d$. Also, for a generic $d$-dimensional vector $k$ we define $k_{-|\mu|}\equiv- k_{|\mu|}$.

A NBW of length $n\geq 1$ in $\bbZ^d$ is a sequence $\omega=(x_0,x_1,\dots,x_n)$, $x_i\in \bbZ^d$, such that $\lVert x_{i+1}-x_i\rVert = 1$ and $x_{i+2}\neq x_i$. We call $\alpha$ the \emph{start direction} of $w$  if $x_1-x_0=e_{\alpha}$. We call $\mu$ the \emph{end direction} of  $w$  if $x_n-x_{n-1}=e_{\mu}$.

Without loss of generality, we shall assume $x_0=0$. 
Let us call $a^{\alpha}_n(x)$ the number of NBW of length $n$ in $\mathbb{Z}^d$ starting from the origin in the direction $\alpha$ and ending in $x$ (i.e. $x_n=x$).
It is also convenient to define $b^{\mu,\alpha}_n(x)$ as the number of NBW of length $n$ in $\mathbb{Z}^d$ starting from the origin in the direction $\alpha$ and ending in $x$ whose end direction is \emph{not} $\mu$. These definitions lead to the following relations:
\begin{align}
\label{bx}
b^{\mu,\alpha}_n(x) &= \sum_{\nu\neq \mu} b^{-\nu,\alpha}_{n-1}(x-e_\nu) \qquad &n\geq 2\\
b_1^{\mu,\alpha}(x) &=  (1-\delta_{\mu\alpha})\delta(x-e_\alpha) 
\end{align}
and 
\begin{align}
a^{\alpha}_n(x) &= \sum_\nu b^{-\nu,\alpha}_{n-1}(x-e_\nu) \qquad &n\geq 2\\
a^{\alpha}_n(x) &=  b^{\nu,\alpha}_{n}(x) +b^{-\nu,\alpha}_{n-1}(x-e_\nu) \qquad &n\geq 2,\forall \nu\\
\label{ax}
a^{\alpha}_1(x) &=  \delta(x-e_\alpha). 
\end{align}
 
For a generic (well behaved) function $f(x)$, $x\in\bbZ^d$, we shall denote its Fourier transform with $\hat{f}(k)$, $k\in [-\pi,\pi]^d$, so that
\begin{align}
\hat{f}(k) &= \sum_{x\in \mathbb{Z}^d} f(x)\, e^{i k x} \\
f(x) &= \int_{[-\pi,\pi]^d} \frac{\dd k}{(2\pi)^d}\ \hat{f}(k)\, e^{i k x} \ .
\end{align}
In Fourier space Equations (\ref{bx} - \ref{ax}) read
\begin{align}
\label{bk}
\hat b^{\mu,\alpha}_n(k) &= \sum_{\nu\neq \mu} \hat b^{-\nu,\alpha}_{n-1}(k)\, e^{ik_\nu} \qquad &n\geq 2\\
\hat b_1^{\mu,\alpha}(k) &=  (1-\delta_{\mu\alpha})\,e^{ik_\alpha}
\end{align}
and 
\begin{align}
\label{ak1}
\hat a^{\alpha}_n(k) &= \sum_\nu \hat b^{-\nu,\alpha}_{n-1}(k)\,e^{ik_\nu} \qquad &n\geq 2\\
\label{ak2}
\hat a^{\alpha}_n(k) &=  \hat b^{\nu,\alpha}_{n}(k) + \hat b^{-\nu,\alpha}_{n-1}(k)\, e^{ik_\nu} \qquad &n\geq 2,\forall \nu\\
\hat a^{\alpha}_1(k) &=  e^{i k_\alpha}. 
\label{ak}
\end{align}
We now denote with $\vec{\hat a}_n(k)$ the $2d$-dimensional vector with elements $\hat a^{\alpha}_n(k)$, and with $ \hat{\mathbf{b}}_n(k)$ the $2d\times2d$ matrix with elements $\hat b_n^{\mu,\alpha}(k)$. We also indicate with $\vec{1}$  the all ones vector and with $\mathbf{1}$ the all ones matrix, with $\mathbf{I}$ the matrix identity, with $\vec{v}^{\,T}$ vector transposition, and introduce the matrices $\mathbf{D}(k)$ and $\mathbf{J}$ with elements:
\begin{align}
D^{\mu\alpha}(k) &= \delta_{\mu,\alpha}\, e^{i k_\mu}, \\
J^{\mu\alpha} &= \delta_{\mu,-\alpha} \ .
\end{align}
We can then compactly write Eqs.  (\ref{ak1} - \ref{ak2}) as
\begin{align}
\vec{\hat a}_n^{\,T}(k) =\vec{1}^{\,T} \mathbf{D}(-k) \hat{\mathbf{b}}_{n-1}(k)  \\
\vec{1}\vec{\hat a}_n^{\,T}(k) =\hat{\mathbf{b}}_{n}(k) + \mathbf{D}(k) \mathbf{J} \hat{\mathbf{b}}_{n-1}(k)  
\end{align}
These recurrent relations can be conveniently solved in terms of generating functions. For a succession $f_n, n\geq 1$, we formally define its generating function as:
\begin{equation}
F_z = \sum_{n=1}^{+\infty} f_n\, z^n.
\end{equation}
Therefore, associating  $\vec{A}_z(k)$ to $\vec{\hat{a}}_n(k)$ and $\hat{\mathbf{B}}_z(k)$ to   $\hat{\mathbf{b}}_n(k)$, we obtain 
\begin{align}
\label{AK1}
\vec{A}_z^{\,T}(k) &=z\vec{\hat a}_1^{\,T}(k) +z\vec{1}^{\,T} \mathbf{D}(-k) \hat{\mathbf{B}}_{z}(k)  \\
\label{AK2}
\vec{1}\vec{A}_z^{\,T}(k) &=\hat{\mathbf{B}}_{z}(k)+z\vec{1}\vec{\hat a}_1^{\,T}(k)-z\hat{\mathbf{b}}_1(k) + z\mathbf{D}(k) \mathbf{J} \hat{\mathbf{B}}_{z}(k)  .
\end{align}
Last equation leads to 
\begin{equation}
\label{BK}
\hat{\mathbf{B}}_{z}(k) = \left(\mathbf{I}+z\mathbf{D}(k) \mathbf{J})\right)^{-1}\big(\vec{1}\vec{A}_z^{\,T}(k)-z\vec{1}\vec{a}_1^{\,T}(k)+z\mathbf{b}_1(k)\big),
\end{equation}
which, inserted in \eqref{AK2} gives
\begin{align}
\label{AKalmostfin}
\vec{A}_z^{\,T}(k) &=z\frac{\vec{a}_1^{\,T}(k) +z\vec{1}^{\,T} \mathbf{D}(-k) \left(\mathbf{I}+z\mathbf{D}(k) \mathbf{J})\right)^{-1}\big(\mathbf{b}_1(k)-\vec{1}\vec{a}_1^{\,T}(k)\big)}
{1-z\, \vec{1}^{\,T} \mathbf{D}(-k) \left(\mathbf{I}+z\mathbf{D}(k) \mathbf{J})\right)^{-1}\vec{1}}\ .
\end{align}
This expression is particularly convenient since now we can exploit the relation
\begin{equation}
\left(\mathbf{I}+z\mathbf{D}(k) \mathbf{J})\right)^{-1} = \frac{1}{1-z^2}\left(\mathbf{I}-z\mathbf{D}(k) \mathbf{J}\right) \ ,
\end{equation}
that can be easily derived from the identity $\mathbf{J} \mathbf{D}(k) \mathbf{J} \mathbf{D}(k)=\mathbf{I}$.

Expression \eqref{AKalmostfin} contains a  scalar denominator and no matrix inversions, so it can be easily computed. It reads:
\begin{align}
\label{Afinal}
A_z^{\alpha}(k) &=\frac{z(e^{i k_\alpha}-z)}{1+z^2 (2d-1)-2d z D(k) } \ ,
\end{align}
with 
\begin{equation}
D(k) \equiv \frac{1}{d}\sum_{\mu=1}^d \cos(k_\mu) \ .
\end{equation}
Using this result in Eq. \eqref{BK} we also obtain
\begin{align}
\hat{B}^{\mu,\alpha}_{z}(k) &= \frac{1}{1-z^2}\big((1-z e^{i k_\mu})A^\alpha_z(k)-z(\delta_{\mu,\alpha}\,e^{ik_\mu}-z\delta_{\mu,-\alpha})\big) \ .
\end{align}

From last equation we can derive our final result for the generating function of the number of NBWs with start and end direction $\alpha$ and $\mu$ respectively, which we call $\hat{B}_z(k;\mu,\alpha)$, through a linear transform given by the matrix
\begin{equation}
S_{\mu \nu} \equiv(\mathbf{1}-\mathbf{I})^{-1}_{\mu,\nu}=\frac{1}{2d-1}-\delta_{\mu\nu} \ .
\end{equation}
Therefore the final result is 
\begin{equation}
\label{Bfinal}
\hat{B}_z(k;\mu,\alpha)=(S\,\hat{\mathbf{B}}(k))_{\mu\alpha} = \frac{c^{\mu,\alpha}_z(k) }{1+z^2 (2d-1)-2d z D(k) }
\end{equation}
where we defined
\begin{align}
c^{\mu,\alpha}_z(k)&\equiv\frac{z}{1-z^2} \big[z \big((e^{i k_\alpha} - z) (e^{i k_\mu} - z) + \delta_{\mu,-\alpha} (z (2 d D(k) + z - 2 d z)-1)\big) -\delta_{\mu\alpha} e^{i k_\alpha} (2 d (D(k) - z) z + z^2-1)\big] \ .
\end{align}
One can check that $\sum_\mu \hat{B}_z(k;\mu,\alpha)=A_z^{\alpha}(k)$ as given in Eq. \eqref{Afinal}, as expected. The generating function for the total number of NBWs  $\hat{B}_z(k)\equiv\sum_{\mu,\alpha} \hat{B}_z(k;\mu,\alpha)$, is given by
\begin{equation}
\hat{B}_z(k) = \frac{2dz(D(k)-z)}{1+z^2 (2d-1)-2d z D(k)}
\end{equation}
This expression differs form the one given in Ref. \cite{Fitzner2013} for the different convention on the number of NBW of length zero (1 for them, 0 for us).

Notice that, since $D(k)\sim 1-k^2/2d$  for small $k$, all the generating functions we derived in this Appendix have a pole in $z=1/(2d-1)$ when $k=0$, given by the denominator of last equation.

\section{The Symmetry Factor} \label{sec:symmetry_fact}

The symmetry factor must be considered when we sum over all possible way of realizing a given fat-diagram on the lattice.
The naive way of doing so would be to attach a label to all vertex in the diagram and to all directions of the lines in the diagram. Then one would sum over the position of the vertexes and over the directions. However what could happen is that there are a different set of assignments that correspond to the same diagram and this would cause an over-counting. 

Formally the problem is to determine the minimal amount of information needed to specify the position of the diagram unambiguously.
This may be less than the position and directions of each vertex and lines in the graph if the graph possesses some symmetry.  For instance, if a graph is symmetric under the interchange of two vertexes, its positions must be specified by an un-ordered couple.

In order to solve this problem, we may label all the internal vertexes and then label the legs of each vertex.
Then we consider a different labeling of the vertexes and of the legs and we may ask if the adjacency matrix of the graph is different from the original one.
The number of different labelings that give the same adjacency matrix is the symmetry group of the graph.

Now it is clear that the symmetry factor of a fat-diagram depends only on its topological structure and not on the actual lengths of its internal lines, in other words, it is the same of the corresponding Feynman diagram.  

In the case of the Feynman diagram the Wick theorem grants that if I multiply by the permutation of the vertexes and by the permutation of the internal lines, this cancels both the overall factorials coming from the Taylor expansion and the factor $1/k!$ that is in the coupling constant.
The result, however, has to be divided by a symmetry factor that is in order to avoid multiple counting. 

According to \cite{Cheng1984, Kaku1993} in a real scalar theory the symmetry factor of a generic diagram can be immediately computed as:
\begin{equation}
S= p 2^{\gamma}\prod_{n=2,3,...} (n!)^{\alpha_n}
\label{symmetry_fact}
\end{equation}
where $\alpha_n$ is the number of pairs of vertices connected by n identical self-conjugated lines, $\gamma$ is the number of lines connecting a vertex to itself and finally, $p$ is the number of permutations of vertices retaining the structure of the diagram unchanged for fixed external lines.
According to Eq. (\ref{symmetry_fact}) we immediately see that in our case the symmetry factors of the diagrams shown in the upper part of Fig. (\ref{fig:fat-diagrams}) are $2$, $2$, $6$ and $2$.
Let us neglect for a while the formula above and focus on an ordinary $\varphi^4$ theory. The symmetry factor of the cactus diagram is given by $S=\frac{4!}{4 \cdot 3}=2$ because one has $4$ possibilities to connect the internal vertex with the leg on the left side and other $3$ possibilities to connected it with the leg on the right side, as one link is already fixed. 
In a similar way, for the Saturn diagram, $S= \frac{2 (4!)^2}{ 4 \cdot 4 \cdot 2 \cdot 3!}=6$, because it has two vertexes each contributing with a factor $4!$, and there are $4$ different ways to attach each external leg to one of the internal vertexes and more $3!$ ways to permute the internal lines with themselves. We implicitly assume that all the lines are distinguishable. However, if the diagram is sensitive to the direction, its multiplicity depends only on the number of ways needed to construct the same topology, once given a constraint on the direction.
Extending this computation to our particular case, the symmetry factors are the same as those derived for a generic field theory.

\bibliography{biblio.bib}

\end{document}